\documentclass{aastex631}
\usepackage{threeparttable} 
\usepackage{graphicx} 
\usepackage{rotating} 
\usepackage{amsmath}
\usepackage{booktabs}
\usepackage[utf8]{inputenc}
\usepackage{booktabs}
\usepackage{multirow}
\usepackage{xcolor}
\usepackage{threeparttable}

\begin{document}

\title{High-Time-Cadence Spectroscopy and Photometry of Stellar Flares on M-dwarf YZ Canis Minoris with Seimei Telescope and TESS. I. Discovery of Rapid and Short-Duration Prominence Eruptions}

\author{Yuto Kajikiya}
\affiliation{Department of Earth and Planetary Sciences, Tokyo Institute of Technology \\
2-12-1 Ookayama, Meguro-ku, Tokyo 152-8551, Japan}

\author[0000-0002-1297-9485]{Kosuke Namekata}
\affiliation{The Hakubi Center for Advanced Research, Kyoto University, Kyoto 606-8302, Japan}
\affiliation{Department of Physics, Kyoto University, Kitashirakawa-Oiwake-cho, Sakyo-ku, Kyoto, 606-8502, Japan}
\affiliation{Division of Science, National Astronomical Observatory of Japan, NINS, Osawa, Mitaka, Tokyo, 181-8588, Japan}

\author[0000-0002-0412-0849]{Yuta Notsu}
\affiliation{Laboratory for Atmospheric and Space Physics, University of Colorado Boulder 3665 Discovery Dr, Boulder, CO 80303, USA}
\affiliation{National Solar Observatory, 3665 Discovery Drive, Boulder, CO 80303, USA}

\author[0000-0003-0332-0811]{Hiroyuki Maehara}
\affiliation{Okayama Branch Office, National Astronomical Observatory of Japan\\ 
3037-5 Honjo, Kamogata-cho, Asakuchi, Okayama 719-0232, Japan}

\author[0000-0001-8033-5633]{Bunei Sato}
\affiliation{Department of Earth and Planetary Sciences, Tokyo Institute of Technology \\
2-12-1 Ookayama, Meguro-ku, Tokyo 152-8551, Japan}

\author[0000-0001-9588-1872]{Daisaku Nogami}
\affiliation{Department of Astronomy, Kyoto University, 
Kitashirakawa Oiwake-cho, Sakyo-ku, Kyoto 606-8502, Japan}

\begin{abstract}
M-dwarfs show frequent flares and associated coronal mass ejections (CMEs) may significantly impact close-in habitable planets. M-dwarf flares sometimes show red/blue asymmetries in the H$\alpha$ line profile, suggesting prominence eruptions as an early stage of CMEs. However, their high-time-cadence observations are limited. We conducted spectroscopic monitoring observations of the active M-dwarf YZ Canis Minoris with $\sim$1 minute time cadence using the Seimei telescope, simultaneously with the optical photometric observations by Transiting Exoplanet Survey Satellite. We detected 27 H$\alpha$ flares with H$\alpha$ energies ranging from 1.7 $\times$ 10$^{29}$ to 3.8 $\times$ 10$^{32}$ erg and durations from 8 to 319 minutes. Among them, we identified 3 blue asymmetry and 5 red asymmetry events based on criteria using the Bayesian Information Criterion. The maximum velocity of the blue- and red-shifted components ranges from 250 to 450 km s$^{-1}$ and 190 to 400 km s$^{-1}$, respectively. The duration and time evolution show variety, and in particular, we discovered rapid, short-duration blue/red asymmetry events with the duration of 6--8 minutes. Among the 8 blue/red asymmetry events, two blue and one red asymmetry events are interpreted as prominence eruptions because of their fast velocity and time evolution. Based on this interpretation, the lower limit of occurrence frequency of prominence eruptions can be estimated to be $\sim$1.1 events per day. Our discovery of short-duration events suggests that previous studies with low time cadence may have missed these events, potentially leading to an underestimation of the occurrence frequency of prominence eruptions/CMEs.

\end{abstract}


\section{Introduction} \label{sec:intro}
Solar and stellar flares are explosive phenomena that occur suddenly in the corona of the Sun and stars \citep[e.g.,][]{Kowalski2024}. They are observed as an increase in electromagnetic radiation, ranging from X-rays to radio waves. These flares are believed to be caused by the conversion of magnetic energy into kinetic and thermal energy through magnetic reconnection \citep{Shibata2011}. Some of the energy released in the corona is transported to the chromosphere and photosphere via heat conduction and high-energy particles. This leads to chromospheric evaporation, chromospheric condensation, and radiation from the corona, transition region, chromosphere, and photosphere (e.g., \citealt{Fisher1985,Allred2006}). Part of the magnetic energy is released for plasma ejections known as prominence/filament eruptions. On the Sun, 
they often develop into coronal mass ejections (CMEs), which can significantly impact Earth's environment (e.g., \citealt{Bisi2010}; \citealt{Gopalswamy2016}).

In recent years, M-dwarfs are considered as prime targets for the search for habitable planets \citep{Nutzman2008,Gilbert2020} and have been observed to show very high flare activity. Superflares with more than ten times the energy of the largest solar flares are observed \citep{Namizaki2023}. 
If CMEs associated with these superflares can eject from the stellar surface, the impact on surrounding close-in planets would be significant. Large CMEs could cause atmospheric erosion if they collide with a close-in planet \citep{Lammer2007}. Additionally, high-energy particles accelerated by fast shocks produced during CMEs could promote chemical reactions in planetary atmosphere  \citep{Airapetian2016}. This could potentially lead to the origin of life-essential molecules such as amino acids \citep{Kobayashi2023}. Since M-dwarfs have habitable zones closer to the star compared to G-dwarfs and K-dwarfs, such impacts on close-in habitable planets can be greater \citep{Lingam2018}.
Therefore, understanding the frequency and energy of prominence/filament eruptions, and eventually CMEs, in active M-dwarfs is crucial for evaluation of habitable condition of exoplanets.

Over the past several decades, optical spectroscopic observations have revealed that chromospheric line profiles often show blue asymmetries during flares on M-dwarfs, suggesting possible prominence/filament eruptions (e.g., \citealt{Houdebine1990,Vida2016,Vida2019,Maehara2021,Notsu2024,Inoue2024}).
\cite{Vida2016} reported a blueshifted component with a maximum radial velocity of 675 km s$^{-1}$ during an H$\alpha$ flare on the M4 dwarf V374 Peg. Additionally, \cite{Vida2016} observed an H$\alpha$ flare where a red asymmetry followed the blue asymmetry in the H$\alpha$ line, suggesting that some of the ejected low-temperature plasma returns to the stellar surface. \cite{Honda2018} reported a long-duration blue asymmetry lasting about 2 hours at a velocity of approximately 100 km s$^{-1}$ during an H$\alpha$ flare on the M4.5 dwarf EV Lac. \cite{Maehara2021} reported a blue asymmetry lasting about 1 hour without significant white-light brightening (a possible non-white-light flare) during an H$\alpha$ flare on the M4.5 dwarf YZ CMi. Furthermore, \cite{Inoue2024} reported a blue asymmetry appearing about 1 hour after the peak of an H$\alpha$ flare on the M4.5 dwarf EV Lac. \cite{Notsu2024} reported 7 blue asymmetries among 41 H$\alpha$ flares in the M-dwarfs YZ CMi, EV Lac, and AD Leo, with durations ranging from approximately 20 minutes to 150 minutes and velocities ranging from about 73 km s$^{-1}$ to 122 km s$^{-1}$. It has been discussed that they can be interpreted as prominence/filament eruptions based on the velocity and mass. However, the time cadence ($>$5 min) of the previous observations is not necessarily high enough to capture and trace the rapid velocity changes of prominence eruptions on M-dwarfs with strong surface gravity. Actually, some solar prominence/filament eruptions change their velocities in less than 5 minutes \citep{Otsu2022}. Higher-time-cadence ($\sim$1 min) is essentially required not to miss even such short-duration/fast phenomena especially for M-dwarfs.

Meanwhile, the nature and origin of the H$\alpha$ blue asymmetries during M-dwarf flares are still under debate, and simultaneous observations with multiple wavelength ranges and different methods (e.g., photometry) are useful for this purpose. For example, the blue asymmetries could also be explained by the upward flow of low-temperature plasma associated with chromospheric evaporation, as observed in the early stage of solar flares \citep{Tei2018,Canfield1990}. However, white-light flares represent the emission only from flare ribbons originated from chromospheric evaporation/condensation during impulsive phase \citep{Watanabe2013,Namekata2017}, thus it is crucial to compare the emission energy and temporal variation of white-light emission with those of H$\alpha$ blue asymmetries.
One-month continuous photometric observations by the Transiting Exoplanet Survey Satellite (TESS, \citealt{2015JATIS...1a4003R}) have enabled simultaneous photometric and spectroscopic observations, providing a good opportunity to investigate statistical relationships between blue asymmetries and white-light flares, rotational phase which reflects starspots distribution \citep{Ikuta2023}. From these statistical investigations, we can explore the typical properties of blue asymmetries on M-dwarfs, including the location of occurrence.
Previous studies with such simultaneous observations using white-light photometry and H$\alpha$ line spectroscopy \citep{Maehara2021, Namizaki2023, Notsu2024} reported only three blue asymmetries during the non-white-light flares or a partial white-light flare (occurring not concurrently with the white-light flare, but shortly after its decay) observed by TESS. The number of such observations is still small, and there are no systematic studies using monitoring observations by TESS.

In this study, we conducted high-time-resolution (approximately 1 minute) H$\alpha$ line spectroscopic observations of the active M-dwarf YZ CMi using the Seimei telescope \citep{Kurita2020}. 
Simultaneously, high-precision optical photometric observations were conducted by TESS. 
We performed this simultaneous campaign observation for 12 nights. This significantly increases the sample size of observations of H$\alpha$ red/blue asymmetries for white-light/non-white-light flares, which is expected to be helpful to identify the origin of blue asymmetries. For this statistical sample, we incorporated a new criteria to identify the asymmetric or symmetric profiles by using the Bayesian Information Criterion (BIC; Section \ref{subsec:Asymmetry_detection}).
The data and observations are described in Section \ref{sec:Observation and Data}. The analysis methods are explained in Section \ref{sec:Analysis}. The analysis results are presented in Section \ref{sec:Results}. A detailed discussion of the H$\alpha$ blue asymmetries is in Section \ref{sec:Discussion}. The summary and conclusions are provided in Section \ref{sec:Summary and Conclusions}.

\section{Observation and Data} \label{sec:Observation and Data}

\subsection{Target Star: YZ CMi}
YZ CMi is an M4.5-type dwarf known as a flare star, with a thick convective zone and rapid rotation. Zeeman-Doppler Imaging by \cite{Morin2008,Lang2014} reported YZ CMi had a large-scale dipole magnetic field in 2007 and 2008. Table \ref{tab:Stellar Parameters} lists the stellar parameters of YZ CMi. Since the first observation of flares on YZ CMi by \cite{vanMaanen1945}, numerous flares across a wide range of wavelengths from X-rays to radio have been observed. Furthermore, superflares with the energies of {$>$}10$^{33}$ erg, which are more than 10 times larger than the largest solar flares ($\sim$10$^{32}$ erg), have also been detected \citep{Maehara2021}, indicating an exceptionally high level of flare activity.
\begin{table}[ht]
\centering
\begin{threeparttable}
\caption{Stellar Parameters}
\label{tab:Stellar Parameters}
\begin{tabular}{cc}
\hline\hline 
Parameter & Value \\
\hline
Effective Temperature $T_{\text{eff}}$ (K) & $3280 \pm 73$\tnote{†} \\
Rotation Period $P_{\text{rot}}$ (day) & $2.774 \pm 0.0014$\tnote{‡} \\
Stellar Radius $R_{\text{star}}$ ($R_{\odot}$) & $0.37^{+0.03}_{-0.06}$\tnote{§} \\
Stellar Mass $M_{\text{star}}$ ($M_{\odot}$) & $0.37 \pm 0.01$\tnote{*} \\
Distance $d$ (pc) & $5.99 \pm 0.0012$\tnote{**} \\
Radial Velocity $v_{\text{RV}}$ (km s$^{-1}$) & $26.66 \pm 0.0033$\tnote{††} \\
\hline
\end{tabular}
\begin{tablenotes}
\item[†] \cite{Gaidos2014}.
\item[‡] \cite{Maehara2021}.
\item[§] \cite{Baroch2020}.
\item[*] \cite{Cifuentes2020}.
\item[**] \cite{GaiaCollaboration2023}.
\item[††] \cite{Fouqu2018}.
\end{tablenotes}
\end{threeparttable}
\end{table}

\subsection{TESS}
TESS is a space telescope designed for the search of exoplanets using the transit method. It divides the entire sky into sectors of 24° by 96°, conducting continuous photometric observations for approximately 27.4 days per sector. 
TESS covers wavelength range from 6000 to 10000 {\AA}. 
YZ CMi was observed by TESS during observations of Sector 34 from January 14 to February 8, 2021 (Barycentric Julian Date (BJD): 2459229.09 -- 2459245.07). 
In this study, we used the Pre-search Data Conditioned Simple Aperture Photometry (PDC-SAP) light curve obtained from the Mikulski Archive for Space Telescopes (MAST) portal\footnote{https://mast.stsci.edu/}. 
Figure \ref{f:tess_seimei_lightcurve} (a) shows the light curve of YZ CMi observed by TESS at a 20-sec cadence.

\subsection{3.8m Seimei Telescope}
\begin{figure}
    \centering
    \includegraphics[width=1\linewidth]{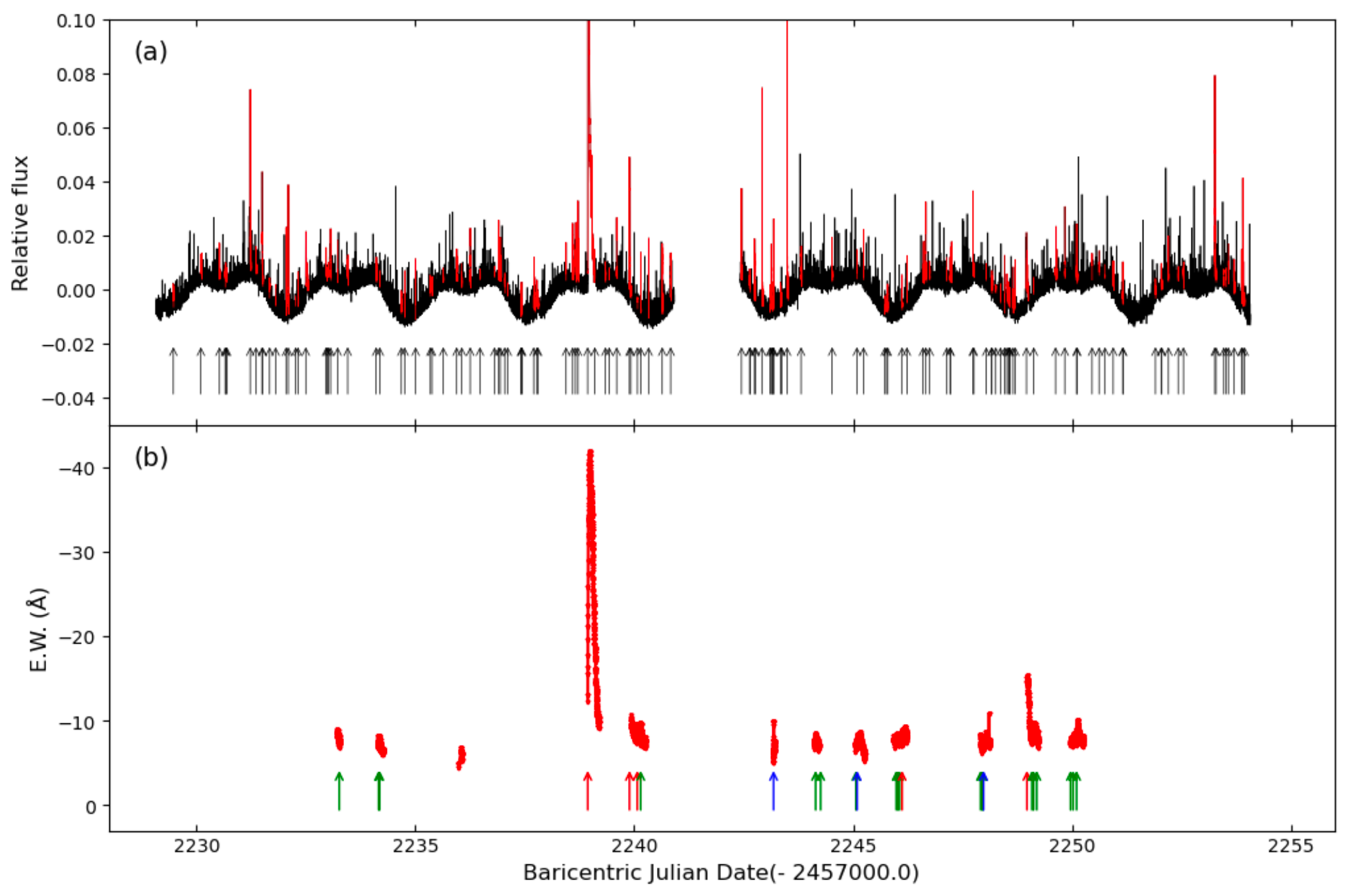}
    \caption{(a) Light curve observed by TESS at a 20-second cadence. The horizontal axis represents BJD, while the vertical axis represents the relative flux normalized to the average ($\delta F/F_{\text{av}}$). The upward arrows in the figure indicate the start times of the detected white-light flares, and the red solid lines show their durations. (b) Light curve of the H$\alpha$ line observed by the Seimei Telescope. The horizontal axis represents BJD, while the vertical axis represents the equivalent width of the H$\alpha$ line. The upward arrows in the figure indicate the start times of the detected H$\alpha$ flares. Red arrows point to flares showing red asymmetry, blue arrows to flares showing blue asymmetry, and green arrows to flares showing no asymmetry.}
    \label{f:tess_seimei_lightcurve}
\end{figure}

The spectroscopic observations were conducted with the Seimei telescope at the same time as the photometric observations at TESS. The Seimei telescope, located at the Okayama Observatory, is a 3.8-meter telescope \citep{Kurita2020} equipped with the low-dispersion spectrograph KOOLS-IFU (Kyoto Okayama Optical Low-dispersion Spectrograph with an optical-fiber integral field unit; \citealt{Matsubayashi2019}). 
The data of YZ CMi in this study were obtained from January 18 to February 4, 2021, using the KOOLS-IFU spectrograph and the VPH683 grism (wavelength range 5800--8000{\AA}, spectral resolution $\lambda /\Delta\lambda\sim 2000$). Observations were conducted over 13 clear nights, with 67 hours of actual observing time.
The observed wavelength range included the H$\alpha$ line (6562.8{\AA}). The velocity resolution around the H$\alpha$ line was approximately 150 km/s, with a sampling of approximately 50 km/s per pixel. 
This line is a crucial chromospheric line indicating flare activity.
The exposure time was set to 60 seconds and the readout time is $\sim$17 seconds, resulting in the time cadence of $\sim$77 seconds.
This high temporal resolution spectroscopy enables detailed investigations of spectral changes in the H$\alpha$ line during stellar flares.
Initial data processing followed the methods outlined by \cite{Namekata2020,Namekata2022,Namizaki2023}, using packages from IRAF \citep{IRAF1986} and PyRAF \citep{Pyraf2012}. 
The process includes the bias subtraction, gain correction, and wavelength calibration.
We did not use the fibers which include the significant cosmic-ray contamination around the H$\alpha$ line.
Additionally, each spectrum was corrected for the radial velocity $v_{\text{RV}}$ of YZ CMi (26.66 km s$^{-1}$; \citealt{Fouqu2018}). The spectra were normalized based on the continuum near the H$\alpha$ line, and the intensity of the H$\alpha$ line was integrated over the range of 6562.8$\pm$15 {\AA},to calculate the equivalent width of the H$\alpha$ line. Figure \ref{f:tess_seimei_lightcurve} (b) shows the light curve of the H$\alpha$ line equivalent width observed by the Seimei Telescope.

\section{Analysis}\label{sec:Analysis}
\subsection{TESS White-Light Flare Detection}\label{subsec:white-light flare detection}
We detected white-light flares from the TESS light curve using the method similar to that of \citet{Maehara2021}.
We analyzed the PDC-SAP light curve of YZ CMi from TESS Sector 34 with the 20s cadence.
Before the flare detection process, we first removed non-flare signals such as long-term trends and brightness variations due to rotation from the light curve. Initially, we fitted the long-term variations in the light curve with a 5th-degree spline function and subtracted the result from the original light curve to eliminate non-flare signals. In the processed light curve, any increase in brightness exceeding 3$\sigma$ ($\sigma$: the average photon noise of the light curve) for three consecutive data points was selected as a flare (cf., \citealt{Chang2015}). The bolometric energy of the detected white-light flares was calculated assuming a blackbody radiation of 10000 K and the TESS response function, following the method described by \cite{Shibayama2013}. The duration of the detected flares was calculated from the point where the brightness increase first exceeded 1$\sigma$ until it decreased below 1$\sigma$ for two consecutive points.

\subsection{H$\alpha$ Flare Detection}\label{subsec:Hα flare detection}

The detection of H$\alpha$ flares was conducted using the light curve of the H$\alpha$ line at the same time as the photometric observations at TESS. 
For the detection method, an increase in the absolute value of the equivalent width of the H$\alpha$ line by more than 3$\sigma$ ($\sigma$: standard deviation of the equivalent width values in the quiescent state) from the quiescent state spectra\footnote{The quiescent state spectra are the averaged spectra of the quiescent state. The quiescent state is defined as the state immediately before any significant increase in brightness was observed. If data from this pre-flare quiescent state were missing, the post-flare state was used as the quiescent state.}, persisting for three consecutive data points, was identified as a flare. The duration over which the absolute value of the equivalent width remained above 1$\sigma$ from the quiescent state spectra was considered as part of the same flare. This method has the uncertainty of potentially counting multiple flares as a single flare or counting a single flare as multiple flares.

Next, we calculate the energy of detected H$\alpha$ flares, following the approach by \cite{Namizaki2023}. First, the H$\alpha$ flux, $F_{\text{H}\alpha}$, was calculated by multiplying the H$\alpha$ equivalent width by the continuum flux near the H$\alpha$ line. The quiescent state continuum flux was derived from values obtained from quiescent state spectra that had been calibrated for flux according to \cite{Kowalski2013}. 
Using $F_{\text{H}\alpha}$ and the distance between YZ CMi and Earth ($d=5.99$ pc; \citealt{GaiaCollaboration2023}), the H$\alpha$ luminosity was then calculated based on the equation:
\begin{equation}
L_{\text{H}\alpha}=4\pi d^2 F_{\text{H}\alpha}
\end{equation}
Following \cite{Notsu2024}, we used TESS's relative flux $\delta$F/F$_{\text{av}}$ to account for the increase in the continuum flux near the H$\alpha$ line during the flare duration. We note that the wavelengths of the TESS band and the H$\alpha$ line are different. Therefore, the changes in the continuum flux near the H$\alpha$ line may be slightly different from the changes in the TESS band flux. Subsequently, the differential luminosity, $L_{\text{H}\alpha,\text{flare}}$ (= $L_{\text{H}\alpha}$ $-$ $L_{\text{H}\alpha,\text{quie}}$, where $L_{\text{H}\alpha,\text{quie}}$ is the quiescent state luminosity), was integrated over the duration of the H$\alpha$ flare to calculate the energy of the H$\alpha$ flare according to the equation:
\begin{equation}
E_{\text{H}\alpha} = \int L_{\text{H}\alpha,\text{flare}}(t) \, dt
\end{equation}
The duration of the detected H$\alpha$ flares was determined from the point where the equivalent width of the H$\alpha$ line increased more than 1$\sigma$ from the quiescent state to the point where it decreased below 1$\sigma$.

\subsection{Detection of H$\alpha$ Line Profile Asymmetry}\label{subsec:Asymmetry_detection}

\begin{figure}
    \centering
    \includegraphics[width=1\linewidth]{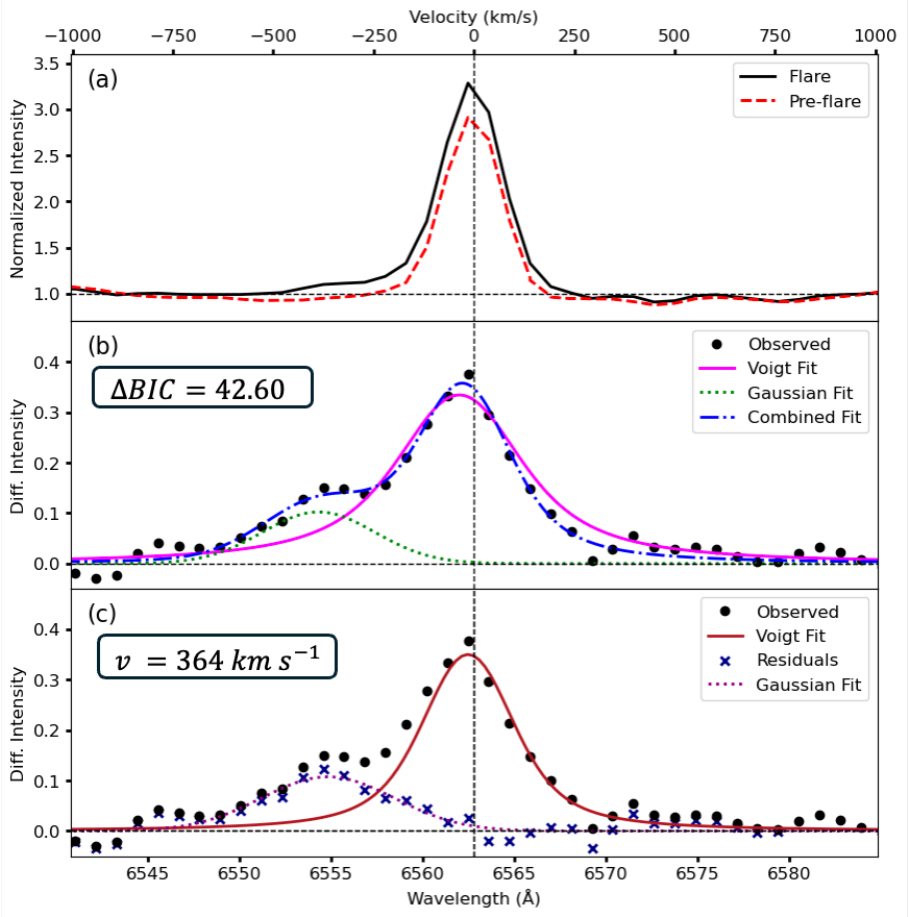}
    \caption{Example of normalized spectra and differential spectra. The lower axis represents the wavelength, and the upper axis shows the Doppler velocity relative to the H$\alpha$ line (6562.8Å). The vertical dashed lines indicate the center wavelength of the H$\alpha$ line. (a) Example of normalized spectra. The black solid line shows the H$\alpha$ line spectrum normalized by the continuum spectrum during the flare period. The red dashed line shows the H$\alpha$ line spectrum normalized by the continuum spectrum in the pre-flare period. 
    (b) Example of differential spectrum fitting for asymmetry determination. 
    Black dots represent observational data and 
    the blue dash-dotted line represents the fit by the combination of Voigt (magenta solid line) and Gaussian (green dotted line)functions, 
    The value in the top left indicates the $\Delta$BIC value for this differential spectrum. (c) Example of differential spectrum fitting for the extraction of blueshifted components. Black dots represent the observational data. The brown solid line represents the Voigt function fitted to the components at longer wavelengths than the H$\alpha$ line center. The navy crosses represent the residuals of the Voigt function (brown solid line) from the observational data (black dots). The dashed black lines represent the Gaussian functions fitted to these residuals (navy crosses).}
    \label{f:Asymmetry_judge}
\end{figure}

We introduce the method for determining the asymmetry of the H$\alpha$ line spectrum. First, to extract the flare components, the normalized spectrum in the quiescent state was subtracted from the normalized spectrum during the flare duration (see Figure \ref{f:Asymmetry_judge} (a)). Next, the differential spectrum created in this section was fitted with two models suitable for symmetric and asymmetric spectra, respectively: a Voigt function (magenta solid line in Figure \ref{f:Asymmetry_judge} (b)) for symmetric spectra and a combined model (a sum of a Voigt function and a Gaussian function, blue dash-dotted line in Figure \ref{f:Asymmetry_judge} (b)) for asymmetric spectra. In this fitting, the center of the Voigt function was fixed within the range of 6562 {\AA} to 6564 {\AA} for both models. This is done to prevent the influence of line profile asymmetry due to the shift between the center of the H$\alpha$ line (6562.8 {\AA}) and the peak of the observed data caused by insufficient wavelength resolution. Additionally, the width $\sigma$ of the Gaussian function was set to be at least half the wavelength resolution during the fitting. Subsequently, BIC for each model was calculated, and the difference in BIC ($\Delta \text{BIC}$) was calculated by subtracting the BIC value of the combined model from the single Voigt model. The BIC is an indicator that evaluates the goodness of a model, taking into account how well the model fits the data and the complexity of the model. The formula for BIC is as follows:
\begin{equation}\label{eq:BIC定義}
\text{BIC} = -2\ln L^* + k\ln n
\end{equation}
where $L^*$ is the maximum likelihood calculated from the likelihood of the observed spectrum and the model function optimized results, $n$ is the number of data points in the spectrum, and $k$ is the number of parameters in the model function. Assuming that the noise for each data point in the spectrum follows a Gaussian distribution with variance $\sigma^2$, the likelihood is given by:

\begin{equation}
L = \prod_{i=1}^{n} \frac{1}{\sqrt{2\pi\sigma^2}} e^{-\frac{(y_i - f(x_i))^2}{2\sigma^2}}
\end{equation}
where $y_i$ is the y-axis value (Diff. Intensity) of the data in the observed spectrum, and $f(x_i)$ is the corresponding y-axis value in the model spectrum (see Figure \ref{f:Asymmetry_judge} (b)). Furthermore, the maximum log-likelihood is calculated as:
\begin{equation}\label{eq:最大対数尤度}
\ln L^* = -\frac{n}{2} \ln(2\pi\sigma^2) - \frac{n}{2}
\end{equation}
\begin{equation}
\sigma = \sqrt{\frac{1}{n} \sum_{i=1}^{n} (y_i - f(x_i))^2}
\end{equation}
The criterion for determining asymmetry was set when $\Delta \text{BIC}$ was greater than 2 for at least two consecutive points. This is because if $\Delta \text{BIC}>0$, the combined model is statistically significantly better than the single Voigt model, and a $\Delta \text{BIC}>2$ indicates a significant difference between the two models \citep{Fabozzi2014}. To exclude influences such as cosmic rays, the criterion was set to require $\Delta \text{BIC}>2$ for at least two consecutive points. If the center of the Gaussian function in the combined model was on the shorter wavelength side of the H$\alpha$ line center, it was determined to be a blue asymmetry. If it was on the longer wavelength side, it was determined to be a red asymmetry.

After the identification of asymmetries, in order to accurately measure the velocity and amplitude with the same method as the previous study, extraction of the redshifted and blueshifted components was conducted by following \citet{Maehara2021,Namizaki2023}.
First, in order to extract the symmetric line center components (referred to as ``flare components''), we fitted a Voigt function to the differential spectrum during the flare duration (brown solid line in Figure \ref{f:Asymmetry_judge} (c)). Unlike the fitting for BIC calculation (see Figure \ref{f:Asymmetry_judge} (b)), the fitting was applied only to the parts on the longer or shorter wavelength side of the H$\alpha$ line center, so that the fitting was not affected from blueshifted or redshifted components, respectively.
Additionally, the center of the Voigt function was fixed at the H$\alpha$ line center. Then, the flare components obtained from the fit was subtracted from the observed data. In this study, the resulting residuals were defined as the `blueshift/redshift components' (navy cross in Figure \ref{f:Asymmetry_judge} (c)). These residuals were then fitted with a Gaussian function (purple dotted line in Figure \ref{f:Asymmetry_judge} (c)), and the Doppler velocity corresponding to the center of the fitted Gaussian function was estimated as the blueshift/redshift velocity. At this time, the width $\sigma$ of the Gaussian function was set to be at least half the wavelength resolution during the fitting. Additionally, the equivalent widths of the central and asymmetric components were calculated by integrating along the wavelength within a ±15 {\AA} range from the center for the Voigt and Gaussian fits, respectively.
Doppler velocity corresponding to the center of the fitted Gaussian function was estimated as the blueshift/redshift velocity.

\clearpage
\begin{sideways} 
    \begin{threeparttable}
    \caption{List of H$\alpha$ Flares} 
    \label{tab:HalphaFlareList}
    \centering
    \begin{tabular}{@{}cccccccccc@{}}
    \hline \hline 
    ID\tnote{a} & Date & Start\_BJD\tnote{b} & $\tau_{\text{H}\alpha}$ (min)\tnote{c} & $E_{\text{H}\alpha}$ (erg)\tnote{d} & $E_{\text{flare}}$ (erg)\tnote{e} & $\tau_{\text{flare}}$ (min)\tnote{f} & Completeness\tnote{g} & WL/NWL\tnote{h} & Asymmetry\tnote{i}  \\
    \hline 
    Y1 & 2021 Jan 18 & 2459233.2613 & 17 & $5.0 \times 10^{29}$ & - & - & complete & NWL & symmetry \\
    Y2 & 2021 Jan 19 & 2459234.1601 & 21 & $8.4 \times 10^{29}$ & - & - & complete & NWL & symmetry \\
    Y3 & 2021 Jan 19 & 2459234.1851 & 37 & $1.3 \times 10^{30}$ & $1.5 \times 10^{31}$ & 9.7 & complete & WL & symmetry \\
    Y4 & 2021 Jan 24 & 2459238.9382 & 319 & $3.8 \times 10^{32}$ & $1.2 \times 10^{34}$ & 235.3 & complete & WL & redshift \\
    Y5 & 2021 Jan 25 & 2459239.9288 & 51 & $4.2 \times 10^{30}$ & $2.5 \times 10^{32}$ & 39.0 & incomplete & WL & redshift \\
    Y6 & 2021 Jan 25 & 2459240.0683 & 36 & $2.3 \times 10^{30}$ & $5.8 \times 10^{30}$ & 2.7 & complete & WL & redshift \\
    Y7 & 2021 Jan 25 & 2459240.1479 & 30 & $1.8 \times 10^{30}$ & $4.0 \times 10^{31}$ & 15.0 & complete & WL & symmetry \\
    Y8 & 2021 Jan 28 & 2459243.1830 & 18 & $2.2 \times 10^{30}$ & $8.2 \times 10^{31}$ & 13.0 & complete & WL & blueshift \\
    Y9 & 2021 Jan 29 & 2459244.1436 & 23 & $9.0 \times 10^{29}$ & - & - & complete & NWL & symmetry \\
    Y10 & 2021 Jan 29 & 2459244.2578 & 17 & $7.2 \times 10^{29}$ & - & - & complete & NWL & symmetry \\
    Y11 & 2021 Jan 30 & 2459245.0639 & 14 & $6.1 \times 10^{29}$ & - & - & complete & NWL & symmetry \\
    Y12 & 2021 Jan 30 & 2459245.0925 & 137 & $9.8 \times 10^{30}$ & - & - & complete & NWL & blueshift \\
    Y13 & 2021 Jan 31 & 2459245.9846 & 49 & $1.6 \times 10^{30}$ & - & - & complete & NWL & symmetry \\
    Y14 & 2021 Jan 31 & 2459246.0258 & 8 & $1.7 \times 10^{29}$ & - & - & complete & NWL & symmetry \\
    Y15 & 2021 Jan 31 & 2459246.0615 & 28 & $6.0 \times 10^{29}$ & - & - & complete & NWL & symmetry \\
    Y16 & 2021 Jan 31 & 2459246.1187 & 146 & $9.2 \times 10^{30}$ & $1.9 \times 10^{31}$ & 8.0 & complete & WL & redshift \\
    Y17 & 2021 Feb 2 & 2459247.9109 & 26 & $1.5 \times 10^{30}$ & - & - & complete & NWL & symmetry \\
    Y18 & 2021 Feb 2 & 2459247.9538 & 10 & $5.9 \times 10^{29}$ & - & - & complete & NWL & symmetry \\
    Y19 & 2021 Feb 2 & 2459247.9681 & 17 & $4.1 \times 10^{29}$ & - & - & complete & NWL & symmetry \\
    Y20 & 2021 Feb 2 & 2459247.9806 & 192 & $1.2 \times 10^{31}$ & - & - & complete & NWL & blueshift \\
    \hline 
    \end{tabular}
    \end{threeparttable}
\end{sideways}
\clearpage 

\clearpage
\begin{sideways} 
    \begin{threeparttable}
    \caption{\textit{(continued)}}
    \centering
    \begin{tabular}{@{}cccccccccc@{}}
    \hline \hline 
    ID\tnote{a} & Date & Start\_BJD\tnote{b} & $\tau_{\text{H}\alpha}$ (min)\tnote{c} & $E_{\text{H}\alpha}$ (erg)\tnote{d} & $E_{\text{flare}}$ (erg)\tnote{e} & $\tau_{\text{flare}}$ (min)\tnote{f} & Completeness\tnote{g} & WL/NWL\tnote{h} & Asymmetry\tnote{i}  \\
    \hline 
    Y21 & 2021 Feb 3 & 2459248.9709 & 122 & $2.8 \times 10^{31}$ & $4.3 \times 10^{32}$ & 72.3 & incomplete & WL & redshift \\
    Y22 & 2021 Feb 3 & 2459249.0835 & 45 & $2.3 \times 10^{30}$ & - & - & complete & NWL & symmetry \\
    Y23 & 2021 Feb 3 & 2459249.1247 & 90 & $5.2 \times 10^{30}$ & $4.7 \times 10^{30}$ & 1.7 & complete & WL & symmetry \\
    Y24 & 2021 Feb 3 & 2459249.1953 & 30 & $1.2 \times 10^{30}$ & - & - & complete & NWL & symmetry \\
    Y25 & 2021 Feb 4 & 2459249.9659 & 61 & $2.0 \times 10^{30}$ & - & - & complete & NWL & symmetry \\
    Y26 & 2021 Feb 4 & 2459250.0240 & 82 & $3.8 \times 10^{30}$ & - & - & complete & NWL & symmetry \\
    Y27 & 2021 Feb 4 & 2459250.1054 & 124 & $7.4 \times 10^{30}$ & $6.6 \times 10^{31}$ & 20.0 & complete & WL & symmetry \\
    \hline 
    \addtocounter{table}{-1}
    \end{tabular}
    \begin{tablenotes}
    \item[a] Label of H$\alpha$ flares
    \item[b] Start time of H$\alpha$ flares (BJD)
    \item[c] Duration of H$\alpha$ flares (min)
    \item[d] Energy of the H$\alpha$ flare (erg)
    \item[e] Bolometric flare energy (erg)
    \item[f] Duration of white-light flares (min)
    \item[g] Completeness. ``complete" indicates a flare that was fully observed from before the flare onset to the quiescent state after the flare, ``incomplete" indicates a flare missing some data during the flare event.
    \item[h] Classification of white-light/non-white-light flares. ``WL" indicates white-light flares detected in Section \ref{subsec:white-light flare detection}, ``NWL" indicates non-white-light flares that were detected in the H$\alpha$ light curve but not in the TESS light curve.
    \item[i] Classification of asymmetry based on Section \ref{subsec:Asymmetry_detection}. ``symmetry" indicates flares with no asymmetry, ``redshift" indicates flares with red asymmetry, and ``blueshift" indicates flares with blue asymmetry.
    \end{tablenotes}
    \end{threeparttable}
\end{sideways}
\clearpage

\section{Results} \label{sec:Results}


\subsection{Detection of H$\alpha$ and white-light flares}\label{subsubsec:HalphaFlareDetectionResults}

Following the methodology described in Section \ref{subsec:Hα flare detection}, a total of 27 H$\alpha$ flares were detected. 
The energy $E_{\rm H \alpha }$ and duration of the detected H$\alpha$ flares $\tau_{\rm H \alpha }$ are listed in Table \ref{tab:HalphaFlareList}. The H$\alpha$ flare energy $E_{\rm H \alpha}$ ranges from 1.7 $\times$ 10$^{29}$ to 3.8 $\times$ 10$^{32}$ erg, and the duration ranges from 8 to 319 minutes, respectively. The total observation time with the Seimei Telescope from January 14 to February 8, 2021, was 2.8 days. The frequency of detectable H$\alpha$ flares with $E_{\rm H \alpha} > 1.0 \times 10^{29}$ erg is approximately 10 events per day.

Using the methods outlined in Section \ref{subsec:white-light flare detection}, 130 white-light flares were detected. The upward arrows in Figure \ref{f:tess_seimei_lightcurve}(a) indicate the start times of the detected white-light flares. The bolometric energy $E_{\text{flare}}$ and the duration $\tau_{\text{flare}}$ of the white-light flares corresponding to the detected H$\alpha$ flares are listed in Table \ref{tab:HalphaFlareList}. The bolometric energy $E_{\text{flare}}$ ranges from 4.7 $\times$ 10$^{30}$ to 1.2 $\times$ 10$^{34}$ erg, and the duration ranges from  1.7 to 235.3 minutes, respectively. The total observation time with TESS from January 14 to February 8, 2021, was 23.5 days. As a result, the frequency of the detectable white-light flares is estimated to be 5.5 events per day. The occurrence frequency of H$\alpha$ flares is higher than that of white-light flares, with only 10 out of 27 H$\alpha$ flares associated with white-light flares. There is not necessarily a one-to-one correspondence between H$\alpha$ and white-light flares. H$\alpha$ flares are not necessarily accompanied by white-light flares. This phenomenon, termed ``non-white-light flare", has been reported in both solar contexts \citep{Watanabe2017} and stellar contexts \citep{Namekata2020}.

\subsection{Determination of H$\alpha$ Line Asymmetry}
\begin{figure}
    \centering
    \includegraphics[width=1\linewidth]{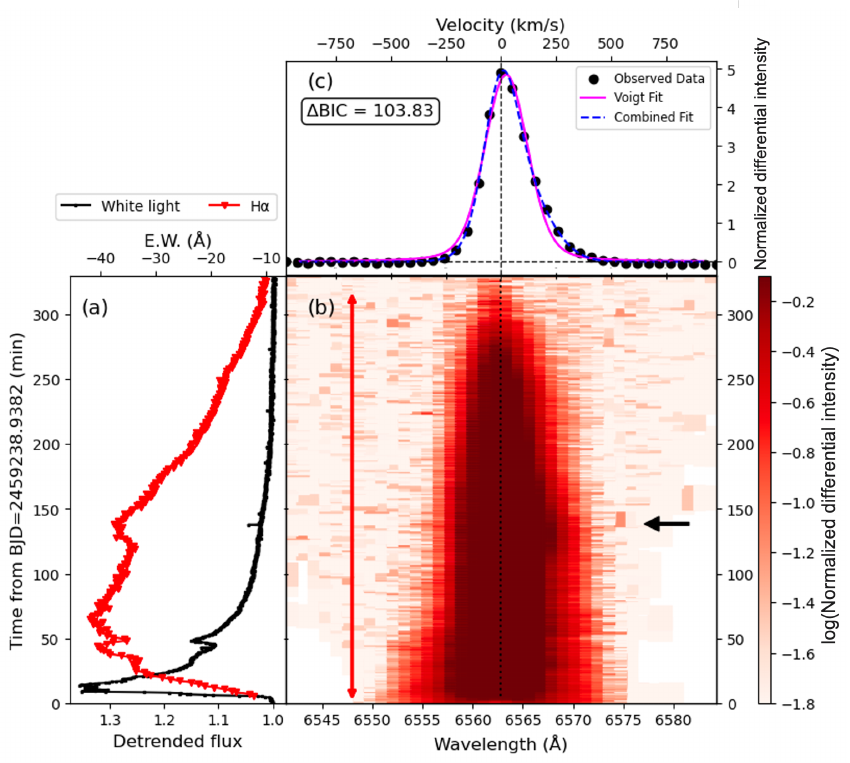}
    \caption{(a) Light curves of white light and H$\alpha$ line for flare Y4. The black solid line represents the detrended light curve of the white light emission showing long-term variations, and the red solid line represents the light curve of the H$\alpha$ line. The left vertical axis represents the detrended flux ($F/F_{\text{star}}$), and the right vertical axis represents the equivalent width of the H$\alpha$ line.(b) Temporal evolution of the H$\alpha$ line profile for Y4. The left vertical axis represents the wavelength, and the right vertical axis shows the Doppler velocity relative to the central wavelength of H$\alpha$ (6562.8\AA). The horizontal axis represents the elapsed time (min) from the start of the flare, with BJD=2459238.9382 as the reference point. The color bar indicates the logarithm of the normalized intensity in the differential spectrum. Black arrows mark the time of maximum $\Delta$BIC, and red bidirectional arrows indicate the duration of red asymmetry.(c) Differential spectrum at the time of maximum $\Delta$BIC. The lower axis represents the wavelength, and the upper axis shows the Doppler velocity relative to the H$\alpha$ line (6562.8\AA). The vertical axis represents the differential intensity, with vertical dotted lines indicating the central wavelength of H$\alpha$, and horizontal dotted lines indicating the zero position of the differential intensity. Black dots represent observational data, the magenta solid line represents a fit by a Voigt function, and the blue dashed line represents a fit by a combination of Voigt and Gaussian functions. The value in the top left indicates the $\Delta$BIC value for this differential spectrum.}
    \label{f:Y4_dynamic_spectrum}
\end{figure}

\begin{figure}
    \centering
    \includegraphics[width=1\linewidth]{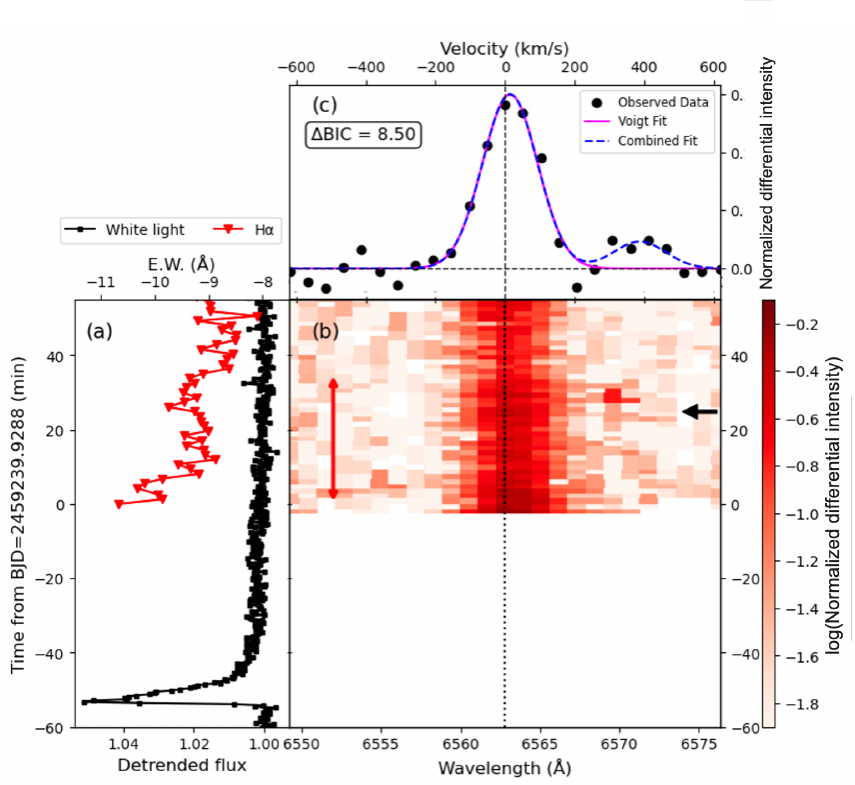}
    \caption{The same as Figure 3, but for the event of Y5.
    }
    \label{f:Y5_dynamic_spectrum}
\end{figure}

\begin{figure}
    \centering
    \includegraphics[width=1\linewidth]{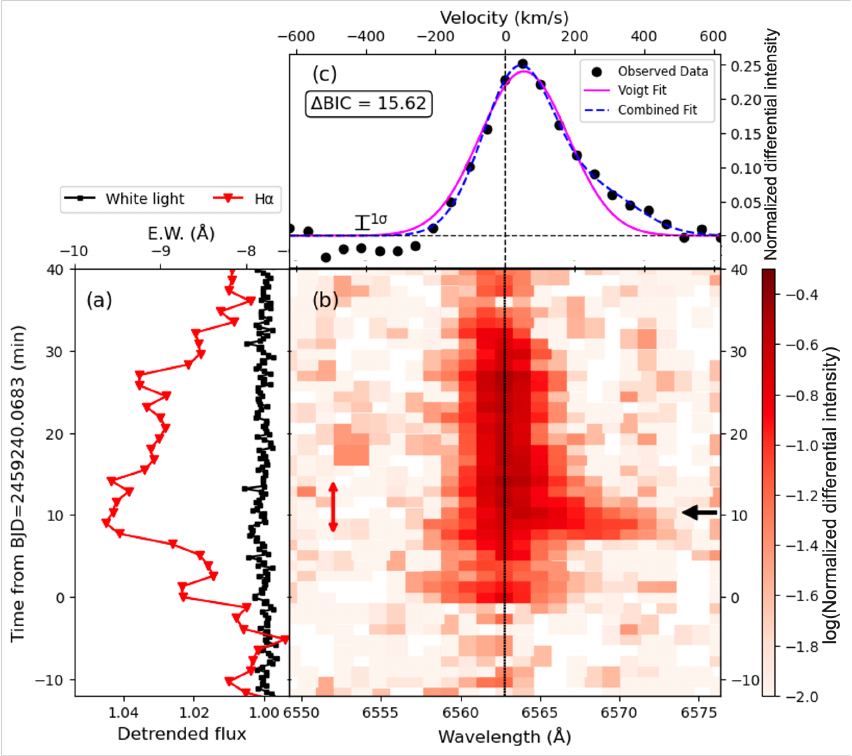}
    \caption{The same as Figure 3, but for the event of Y6. The continuum seems weird on the blue side, but it's within the range of the noise level.}
    
    \label{f:Y6_dynamic_spectrum}
\end{figure}

\begin{figure}
    \centering
    \includegraphics[width=1\linewidth]{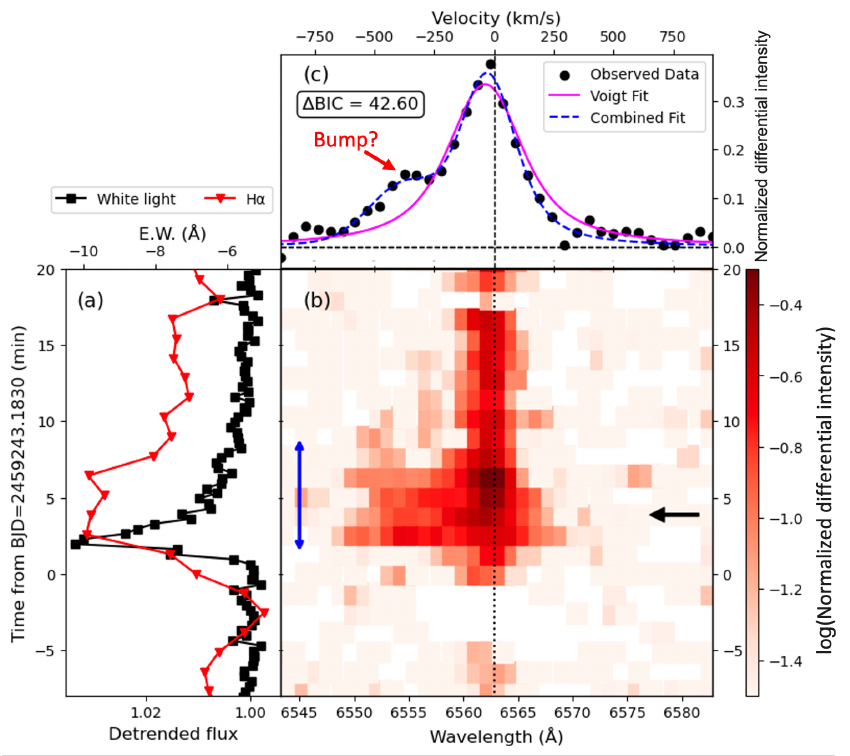}
    \caption{The same as Figure 3, but for the event of Y8. In this case, blue bidirectional arrows indicate the duration of blue asymmetry.
    }
    \label{f:Y8_dynamic_spectrum}
\end{figure}

\begin{figure}
    \centering
    \includegraphics[width=1\linewidth]{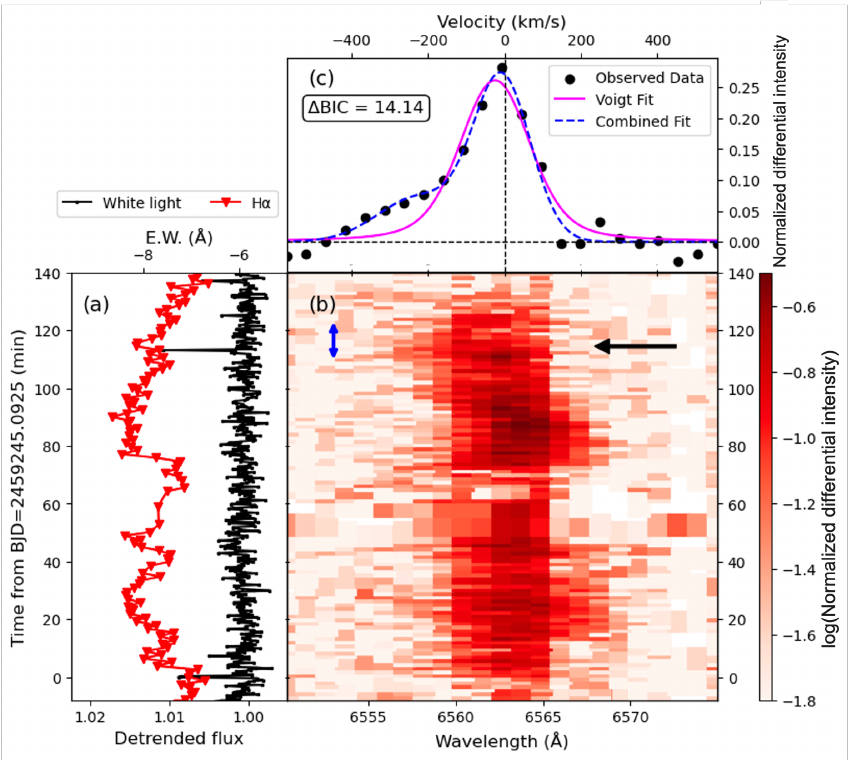}
    \caption{The same as Figure 6, but for the event of Y12.}
    \label{f:Y12_dynamic_spectrum}
\end{figure}

\begin{figure}
    \centering
    \includegraphics[width=1\linewidth]{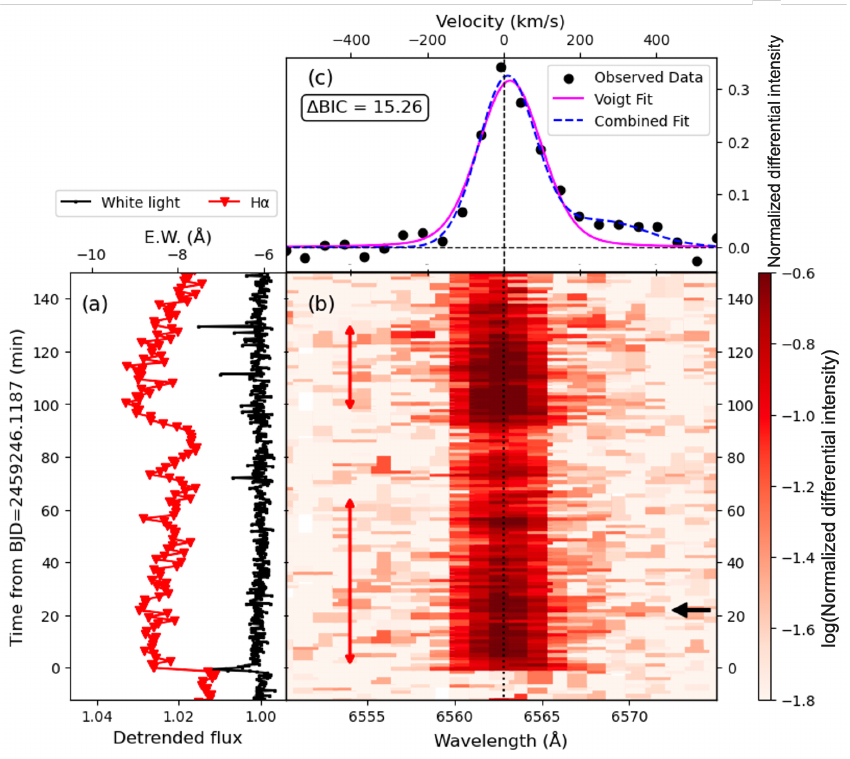}
    \caption{The same as Figure 3, but for the event of Y16.
     }
    \label{f:Y16_dynamic_spectrum}
\end{figure}

\begin{figure}
    \centering
    \includegraphics[width=1\linewidth]{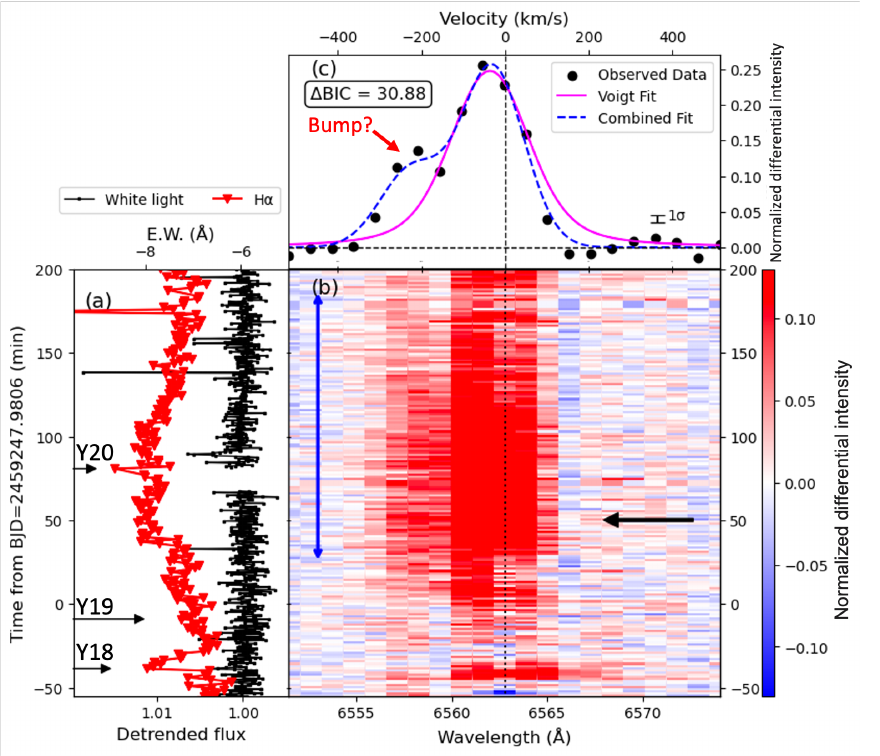}
    \caption{The same as Figure 6, but for the event of Y20. Unlike the others, the color bar indicates the normalized intensity in the differential spectrum, including negative values.
    }
    \label{f:Y20_dynamic_spectrum}
\end{figure}

\begin{figure}
    \centering
    \includegraphics[width=1\linewidth]{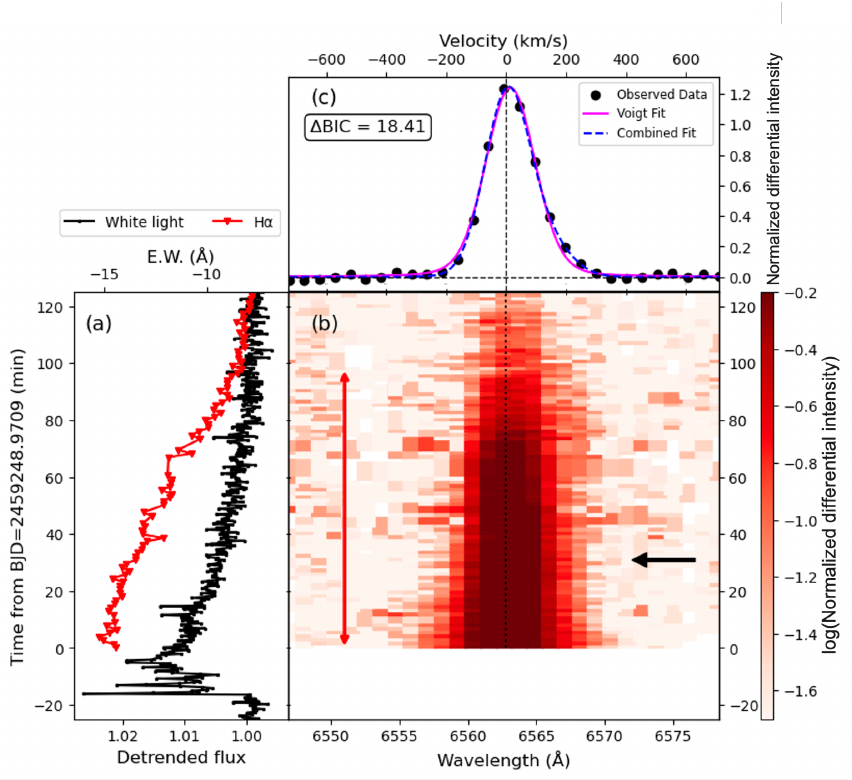}
    \caption{The same as Figure 3, but for the event of Y21.
    }
    \label{f:Y21_dynamic_spectrum}
\end{figure}

Among these 27 H$\alpha$ flares detected in this study, five (Y4, Y5, Y6, Y16, Y21) show red asymmetries, and three (Y8, Y12, Y20) show blue asymmetries, as listed in Table \ref{tab:HalphaFlareList}. The method described in Section \ref{subsec:Asymmetry_detection} was used for the identification of asymmetries.
Figures \ref{f:Y4_dynamic_spectrum} -- \ref{f:Y21_dynamic_spectrum} show the light curves and temporal changes in the H$\alpha$ line profiles for the flares determined to show blue/red asymmetries. 
Specifically, Figure \ref{f:Y4_dynamic_spectrum} corresponds to flare Y4, Figure \ref{f:Y5_dynamic_spectrum} to Y5, Figure \ref{f:Y6_dynamic_spectrum} to Y6, Figure \ref{f:Y16_dynamic_spectrum} to Y16, and Figure \ref{f:Y21_dynamic_spectrum} to Y21, which exhibit red asymmetries. For blue asymmetries, Figure \ref{f:Y8_dynamic_spectrum} corresponds to flare Y8, Figure \ref{f:Y12_dynamic_spectrum} to Y12, and Figure \ref{f:Y20_dynamic_spectrum} to Y20. Other events for which asymmetries were not detected are shown in Appendix \ref{sec:Appendix}.
The maximum velocities $v_{\text{asym,max}}$ and duration of asymmetry $\tau_{\text{asym}}$ for these asymmetric components are shown in Table \ref{tab:Flares showing red/blue asymmetry}. For red asymmetry, $v_{\text{asym,max}}$ ranges from 188 to 400 km s$^{-1}$, and $\tau_{\text{asym}}$ ranges from 6 minutes to 319 minutes. For blue asymmetry, $v_{\text{asym,max}}$ ranges from 202 to 445 km s$^{-1}$, and $\tau_{\text{asym}}$ ranges from 8 minutes to 160 minutes. We evaluated the error of $v_{\text{asym,max}}$ based solely on the Gaussian fitting error. The variation in the observed velocity solution during observations is around 1--2 km s$^{-1}$ and this value is sufficiently small compared to the fitting error as explained in Appendix \ref{Appendix:Velocity Determination Accuracy}. It should be noted that measured velocities $v_{\text{asym,max}}$ are projected ones. The actual values might be higher than that, depending on the configuration. Each blue and redshift event shows very various properties in velocity, duration, and its spectral behavior. Not all of these events are discussed in this paper, but we discuss the blueshift events in the following Section \ref{subsec:OriginsOfBlueAsymmetry}, and one transient redshift event (Y6) in Section \ref{subsec:Y6の起源} in detail.

\begin{table}[ht]
\centering
\resizebox{\textwidth}{!}{%
\begin{threeparttable}
\caption{Flares showing red/blue asymmetry}
\label{tab:Flares showing red/blue asymmetry}
\begin{tabular}{ccccccccc}
\hline\hline
ID\tnote{a} & Asymmetry\tnote{b} & WL/NWL\tnote{c} & $v_{\text{asym,max}}$ (km s$^{-1}$)\tnote{d} & $\tau_{\text{asym}}$ (min)\tnote{e} & $\Delta$BIC$_{\text{max}}$\tnote{f} & $\tau_{H\alpha}$ (min)\tnote{g} & $E_{H\alpha}$ (erg)\tnote{h} & Figure\tnote{i} \\
\hline
Y4 & redshift & WL & $402\pm47$ & 319 & 114.31 & 319 & $3.8 \times 10^{32}$ & Figures 3 \\
Y5 & redshift & WL & $265\pm21$ & 35 & 8.50 & 51 & $4.2 \times 10^{30}$ & Figures 4 \\
Y6 & redshift & WL & $295\pm12$ & 6 & 15.62 & 36 & $2.3 \times 10^{30}$ & Figures 5 \\
Y8 & blueshift & WL & $445\pm36$ & 8 & 42.60 & 18 & $2.2 \times 10^{30}$ & Figures 6 \\
Y12 & blueshift & NWL & $254\pm43$ & 14 & 14.14 & 137 & $9.8 \times 10^{30}$ & Figures 7 \\
Y16$_{\rm{a}}$ & redshift & \multirow{2}{*}{WL} & $242\pm32$ & 66 & 15.26 & \multirow{2}{*}{146} & \multirow{2}{*}{$9.2 \times 10^{30}$} & \multirow{2}{*}{Figures 8} \\
Y16$_{\rm{b}}$ & redshift & & $223\pm21$ & 35 & 5.05 & & & \\
Y20 & blueshift & NWL & $202\pm18$ & 160 & 30.88 & 192 & $1.2 \times 10^{31}$ & Figures 9 \\
Y21 & redshift & WL & $188\pm27$ & 98 & 18.41 & 122 & $2.8 \times 10^{31}$ & Figures 10 \\
\bottomrule
\end{tabular}
\begin{tablenotes}
\item[a] Label of the H$\alpha$ flare
\item[b] Classification of asymmetry based on Section 3.3. ``symmetry" indicates flares with no asymmetry, ``redshift" indicates flares with red asymmetry, and ``blueshift" indicates flares with blue asymmetry.
\item[c] Classification of white-light/non-white-light flares. ``WL" indicates white-light flares detected in Section 3.2, ``NWL" indicates non-white-light flares that were not detected.
\item[d] Maximum velocity of the asymmetrical component (km s$^{-1}$) measured by the fitting process (see Section 3.3).
\item[e] Duration of asymmetry (min)
\item[f] Maximum $\Delta$BIC value
\item[g] Duration of the H$\alpha$ flare (min)
\item[h] Energy of the H$\alpha$ flare (erg)
\item[i] Corresponding Figures showing the light curve and dynamic spectrum.
\end{tablenotes}
\end{threeparttable}%
}
\end{table}


\section{Discussion}\label{sec:Discussion}

\subsection{Discovery and Implication of Rapid, Short-Duration Red/Blue Asymmetries}\label{subsec:高速短時間非対称イベント}

\begin{figure}
    \centering
    \includegraphics[width=1\linewidth]{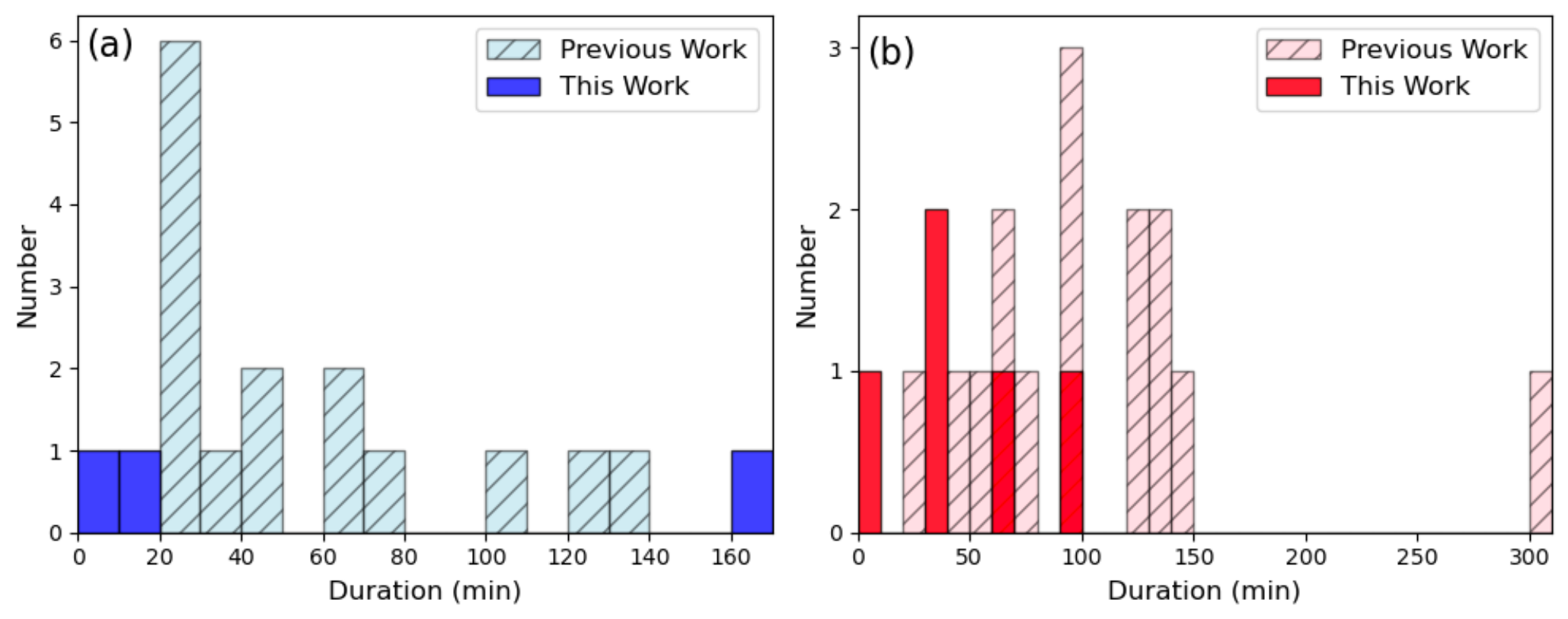}
    \caption{Comparison of the duration of asymmetry during M-dwarf flares with previous studies. (a) Comparison with previous studies on the duration of blue asymmetry. The dark blue bars represent blue asymmetry from this study, while the light blue bars represent blue asymmetry from previous studies. (b) Comparison with previous studies on the duration of red asymmetry. The dark red bars represent red asymmetry from this study, while the light red bars represent red asymmetry from previous studies.}
    \label{f:継続時間統計}
\end{figure}

Since the advantage of this study is the high time cadence observation, we first compare the durations of red/blue asymmetries detected in this study with previous studies. Figure \ref{f:継続時間統計} shows histograms of the duration of (a) blue asymmetries and (b) red asymmetries during M-dwarfs flares detected in this study, as well as those reported in previous studies. Additionally, Table \ref{tab:非対称性継続時間先行研究} lists the events from previous studies displayed in Figure \ref{f:継続時間統計}. From Figure \ref{f:継続時間統計}, previous studies have only reported asymmetries lasting longer than 20 minutes.
We found that the asymmetries during flares Y6 and Y8 in this study are with duration of 6 and 8 minites, respectively.
This is likely due to the $\sim$5-min or longer temporal resolution of observations in previous studies \citep[e.g.,][]{Vida2016,Honda2018,Maehara2021,Wu2022,Wollmann2023,Notsu2024,Inoue2024}, which are not sufficient to detect short-duration asymmetries. 
M-dwarfs have stronger surface gravity compared to G-type or K-type dwarfs, and they could decelerate high-velocity ($\sim$350 km s$^{-1}$) plasma to below 100 km s$^{-1}$ within approximately 5 minutes. In the following sections, we discuss a physical interpretation of the short-duration blueshift in Section \ref{subsec:OriginsOfBlueAsymmetry} and the short-duration redshift in Section \ref{subsec:Y6の起源}.

\begin{table}[ht]
\centering
\begin{threeparttable}
\caption{Summary of asymmetry durations in prior research}
\label{tab:非対称性継続時間先行研究}
\begin{tabular}{cccc}
\hline\hline 
Asymmetry & Duration (min) & Star Name & Reference \\
\hline
blueshift & 24 & V374 Peg & \cite{Vida2016} \\
blueshift & 24 & V374 Peg & \cite{Vida2016} \\
blueshift & 36 & V374 Peg & \cite{Vida2016} \\
blueshift & 120 & EV Lac & \cite{Honda2018} \\
blueshift & 60 & YZ CMi & \cite{Maehara2021} \\
blueshift & 45 & EV Lac & \cite{Inoue2024} \\
blueshift & 20 & YZ CMi & \cite{Notsu2024} \\
blueshift & 20 & YZ CMi & \cite{Notsu2024} \\
blueshift & 100 & YZ CMi & \cite{Notsu2024} \\
blueshift & 60 & YZ CMi & \cite{Notsu2024} \\
blueshift & 48 & YZ CMi & \cite{Notsu2024} \\
blueshift & 70 & EV Lac & \cite{Notsu2024} \\
blueshift & 20 & EV Lac & \cite{Notsu2024} \\
blueshift & 135 & AD Leo & \cite{Notsu2024} \\
\hline
redshift & 24 & V374 Peg & \cite{Vida2016} \\
redshift & 92 & M4-type star\tnote{*} & \cite{Wu2022} \\    
redshift & 300 & YZ CMi & \cite{Namizaki2023} \\
redshift & 47 & AD Leo & \cite{Wollmann2023} \\
redshift & 63 & AD Leo & \cite{Wollmann2023} \\
redshift & 90 & YZ CMi & \cite{Notsu2024} \\
redshift & 120 & YZ CMi & \cite{Notsu2024} \\
redshift & 60 & YZ CMi & \cite{Notsu2024} \\
redshift & 75 & YZ CMi & \cite{Notsu2024} \\
redshift & 90 & YZ CMi & \cite{Notsu2024} \\
redshift & 139 & YZ CMi & \cite{Notsu2024} \\
redshift & 130 & YZ CMi & \cite{Notsu2024} \\
redshift & 140 & YZ CMi & \cite{Notsu2024} \\
redshift & 50 & YZ CMi & \cite{Notsu2024} \\
redshift & 120 & EV Lac & \cite{Notsu2024} \\
redshift & 20 & AD Leo & \cite{Notsu2024} \\ 
\bottomrule
    \end{tabular}
    \begin{tablenotes}
    \item[*] The star name was not mentioned in the paper.
    \end{tablenotes}
    \end{threeparttable}
\end{table}

\subsection{Origins of Blue Asymmetry}\label{subsec:OriginsOfBlueAsymmetry}
\subsubsection{Interpretations of blue asymmetry}
\label{subsubsec:解釈レビュー}
In this section, we discuss the origins of blueshift events. The origins of blue asymmetries can be attributed to three possible phenomena: prominence eruptions \citep[e.g.,][]{Otsu2022}, chromospheric evaporation associated with upflows of cool plasma (chromospheric upflows) \citep[e.g.,][]{Tei2018}, and redshift absorption due to post-flare loops \citep[e.g.,][]{Honda2018,Otsu2022}.
In the following, we discuss these three interpretations based on solar observations and models of M-dwarfs.

First, in solar flares, cool upflows on the chromospheric evaporations are indicated from the blue asymmetry of the chromospheric lines with a velocity of around a few 10 km s$^{-1}$ \citep{Svestka1962,Tei2018,Li2019,Huang2019}.
The typical duration of observable blueshift in chromospheric lines is just a few minutes at a given footpoint \citep{Tei2018,Li2019,Huang2019}, but its total duration could be longer if we see the multiple flare loops in the Sun-as-a-star view.
\cite{Allred2006} performed 1D numerical radiative hydrodynamic simulation for flaring M-dwarf atmosphere and showed that the possible cool upflows with a velocity of $\sim$ 50 km s$^{-1}$ can be formed above the hot evaporation flow as a result of the injection of high energy electron beams into deep chromosphere.
This is thought to be one of the causes of the blue asymmetries in M-dwarf flares.

Second, in solar flares, the downward flow of post-flare loops occurring in the late phase of the flare exhibits red wing absorption in the H$\alpha$ line. The typical velocity of this flow ranges from several tens of km s$^{-1}$ to 100 km s$^{-1}$, and the duration exceeds 60 minutes \citep[e.g.,][]{Liu2013,Song2016,Otsu2022}. \cite{Honda2018} reported a blue asymmetry caused by red wing absorption in the H$\alpha$ line exhibited in the late phase of M-dwarf flares. This suggests the downward flow of post-flare loops.
In M-dwarfs, the background continuum intensity is much cooler than that of the Sun. Therefore, it is unclear whether the post-flare loop can be seen in emission or absorption in the H$\alpha$ line.
While the model presented in \cite{Wollmann2023} suggests that such phenomena are visible in emission, this is also dependent on various parameters and they are not yet fully understood. Therefore, further modeling is necessary.

Third, regarding prominence eruptions, in both the Sun and solar-type stars, these events are observed as H$\alpha$ absorption on the disk and as emission outside the limb \citep{Otsu2022,Namekata2022,Namekata2024}. 
On the other hand, \cite{Leitzinger2022} showed that prominence eruption on the disk of M-dwarfs can also be observed as H$\alpha$ emission using 1D NLTE modeling and cloud model formulation.
Therefore, in M-dwarfs, prominence eruptions on the disk and outside the limb might be observed as H$\alpha$ emission. 
The velocity of prominence eruptions on the Sun ranges from several tens of km s$^{-1}$ to several hundreds of km s$^{-1}$ \citep{Gopalswamy2003,Seki2021}.

These three scenarios might be difficult to distinguish based solely on snapshot spectra. However, the data available for this study were observed with an unprecedented temporal resolution of one minute and are further complemented by comprehensive optical photometric observations from TESS for all events. The TESS white-light flare data, which can have significant constraints on the emission from flare ribbons, can help distinguish flare emissions from other possibilities. Using these high-temporal-resolution photometric and spectroscopic data could enhance our ability to constrain the origins of blue asymmetries. The following sections explore the causes of the three blue asymmetries (Y8, Y12, and Y20), comparing with observations of the Sun and models of M-dwarfs.

\subsubsection{Flare Y8: Rapid, Short-Duration Blue Asymmetry}
\label{subsubsec:Y8の起源}

\begin{figure}
        \centering
        \includegraphics[width=1\linewidth]{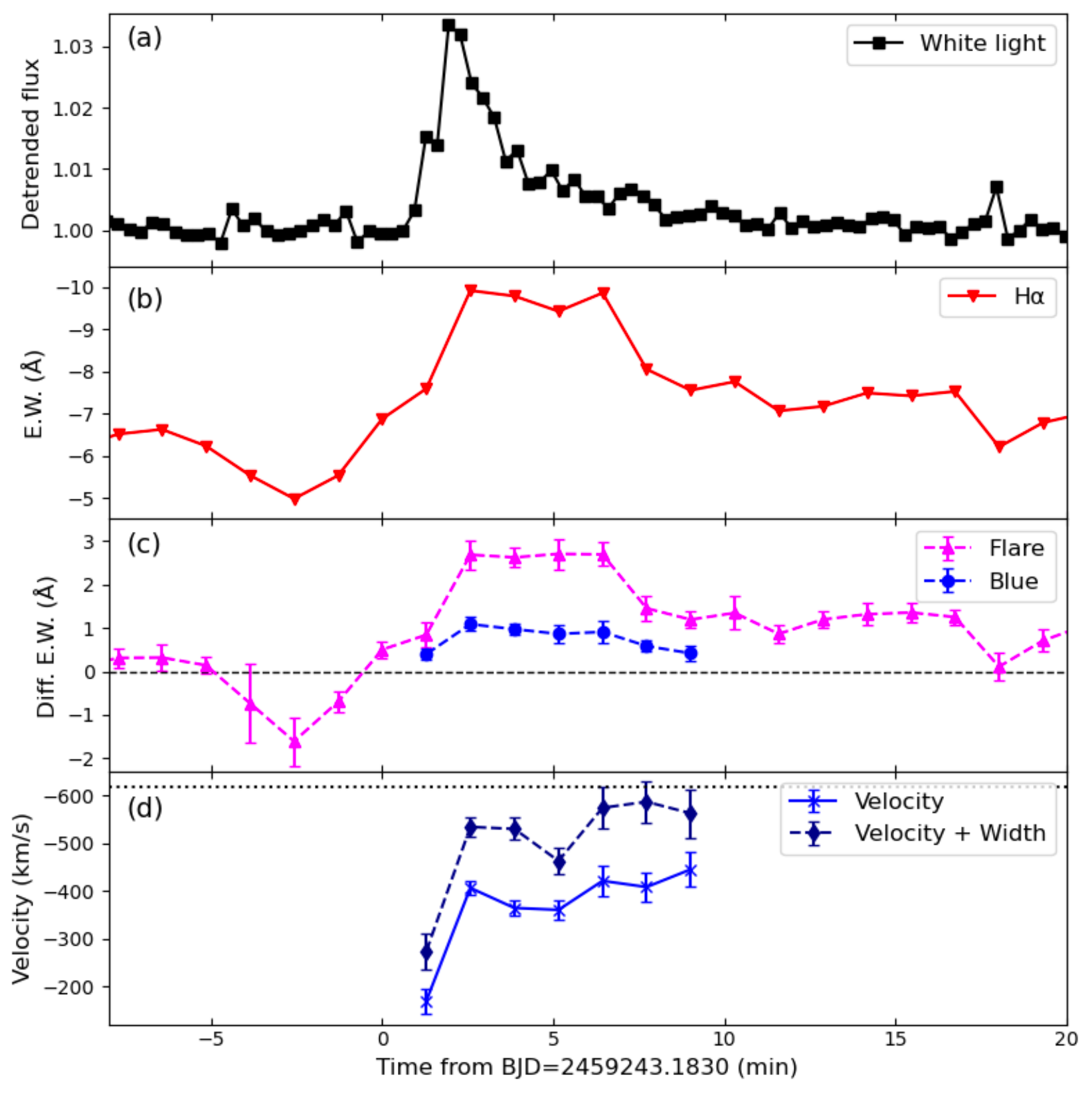}
        \caption{(a) Detrended light curve of flare Y8 The vertical axis represents the detrended flux ($F/F_{\text{star}}$), and the horizontal axis represents the elapsed time (in minutes) from the start of flare Y8, with BJD=2459243.1830 as the reference point.(b) Light curve of the H$\alpha$ line for flare Y8. The vertical axis indicates the equivalent width of H$\alpha$.(c) Time evolution of the equivalent width of the flare component and the blueshifted component for flare Y8. The vertical axis shows the differential equivalent width, with magenta dashed lines indicating the variation in equivalent width of the flare component, and blue dashed lines indicating the variation in equivalent width of the blueshifted component.(d) Velocity changes of the blueshifted component of flare Y8. The vertical axis represents velocity, with a blue solid line showing the velocity changes of the blueshifted component. The navy dashed line shows the evolution of the central velocity plus the Gaussian line width $\sigma$.}
        \label{f:Y8の速度変化}
\end{figure}

In this section, we discuss the origin of the blue asymmetry during the flare Y8 that occurred on 2021 January 28. We first summarize the observational results for the flare Y8. 
Figure \ref{f:Y8_dynamic_spectrum} shows the dynamic spectrum of the Y8 event. 
Also, Figure \ref{f:Y8の速度変化} shows the time evolution of the equivalent width of the flare component and the blueshifted component, and \ref{f:Y8_dynamic_spectrum} (d) shows the time evolution of the velocity of the blueshifted component.
As shown in Table \ref{tab:Flares showing red/blue asymmetry}, the maximum velocity of these blueshifted components was 445 ± 36 km s$^{-1}$, and the duration of asymmetry was 8 minutes.
This velocity is relatively high among typical blueshifted components measured using the same method in time-resolved observations of M-dwarf flares. \citep[$\sim$100 km s$^{-1}$, e.g., ][]{Honda2018,Maehara2021,Notsu2024,Inoue2024}. 
\cite{Vida2019} investigated more than 5500 snapshot spectra of chromospheric lines of M-dwarfs, and reported that 478 spectra among them showed line asymmetries.
The typical maximum velocities for blueshifted components of these spectra are 100--300 km s$^{-1}$, but these maximum velocities were defined as the point where the residual profile merges with the continuum. As a corresponding velocity to be compared with these maximum velocities of 100--300 km s$^{-1}$ in \cite{Vida2019}, the maximum value of the central Gaussian velocity plus the Gaussian line width $\sigma$ of flare Y8 is around $\sim$550 km s$^{-1}$.
Additionally, as discussed in Section \ref{subsec:高速短時間非対称イベント}, the duration of this event is the shortest among H$\alpha$ blueshift events reported in publications.

In addition to the central Gaussian velocity, we here mention the velocity dispersion for each event Y6, Y8, and Y20. As we described above, the observed maximum velocities of these events are below the escape velocity of YZ CMi. However, these blueshifted components are broadened more than velocity resolution $\sim$70 km s$^{-1}$ \citep{Matsubayashi2019}. In Figure \ref{f:Y8の速度変化} and Figure \ref{f:Y12の速度変化} (d), the navy diamonds indicate the evolution of the central velocity plus the Gaussian line width $\sigma$. Similarly, in Figure \ref{f:Y6の速度変化} (d), the brown diamonds show the same evolution. This can be interpreted as the velocity distribution of the expanding prominence \citep{Namekata2024}. Based on this interpretation, we can assume that the width plus central velocity can be regarded as the maximum velocity component of the eruptive prominence. 

Here we discuss the possible origin of this blue asymmetry. 
The typical velocities of chromospheric upflows observed in solar flares (about a few 10 km s$^{-1}$ for solar observation; \citealt{Tei2018}; $\sim$50 km s$^{-1}$ for M-dwarf flare modeling; \citealt{Allred2006}) are much smaller than the maximum velocity of the observed blueshifted components. 
This suggests that chromospheric upflows are unlikely to be the cause. 
Next, Figure \ref{f:Y8_dynamic_spectrum} (c) shows that the line profile exhibits the expanded ``bump-like shape" centered at approximately $-$400 km s$^{-1}$, corresponding to a separate emission peak profile distinct from the H$\alpha$ line center.
This shape cannot be solely explained by redshift absorption from the post-flare loop because the red wing absorption itself cannot make the bump-like feature in the blue wing.
On the other hand, as we described before, prominence eruptions can be observed as blueshifted emissions in M-dwarfs. Therefore, they may make the bump-like feature in the blue wing. The maximum velocity of 445 ± 36 km s$^{-1}$ in this event is relatively fast but almost within the typical velocity of solar prominence eruptions (10--400 km s$^{-1}$; \citealt{Gopalswamy2003,Seki2021}). Both in terms of velocity and profile shape, the characteristics of the prominence are consistent.
Therefore, this blueshifted component is highly likely to indicate a prominence eruption.

Figures \ref{f:Y8の速度変化} (a) and (d) show that this blueshifted component appeared simultaneously with the white-light flare, and the acceleration almost finished before or just after the peak of the white-light flare, followed by almost constant velocity for about 6 minutes.
This is consistent with solar observations where the eruptions are mainly accelerated within the impulsive phase of flares \citep{Aschwanden2021}.
This also supports the possibility of a prominence eruption.

\subsubsection{Flare Y12: Delayed Blue Asymmetry}
\label{subsec:Y12の起源}
\begin{figure}
        \centering
        \includegraphics[width=1\linewidth]{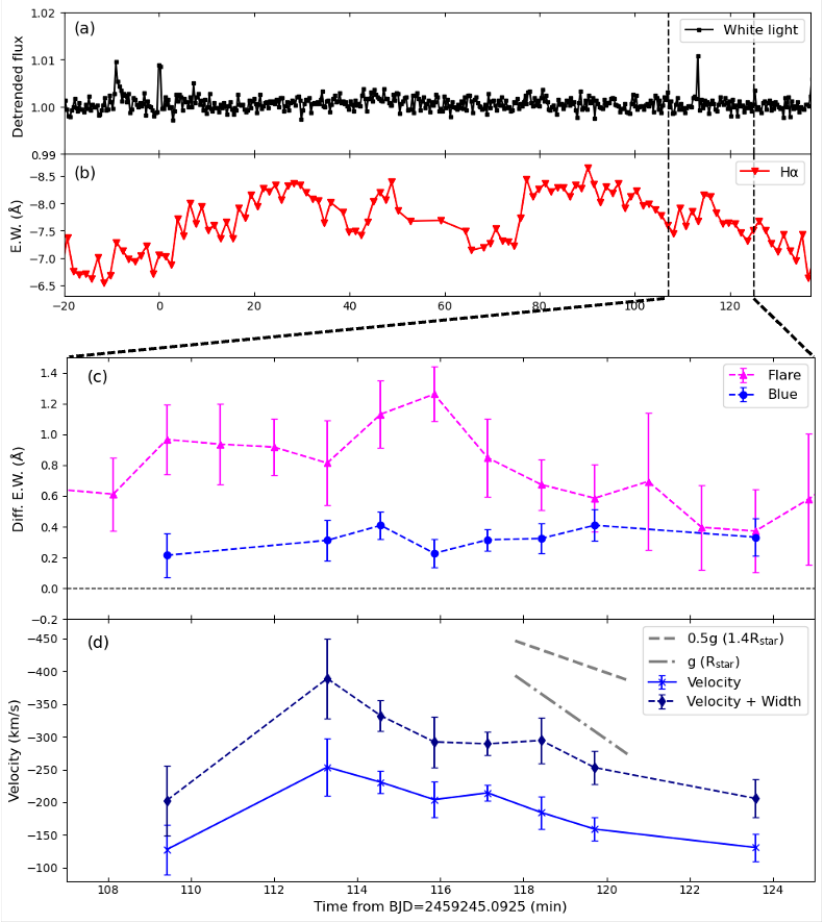}
        \caption{The same as Figure 11, but for the event of Y12. In (d), the gray dot-dashed line indicates the velocity change due to gravitational acceleration at the stellar surface, and the dashed line represents the velocity change due to 0.5 times the gravitational acceleration at the stellar surface, corresponding to the surface gravitational acceleration at 1.4R$_{\text{star}}$.
        }
        
        \label{f:Y12の速度変化}
    \end{figure}

In regard to the blue asymmetry during the flare Y12 on 2021 January 30, Figure \ref{f:Y12_dynamic_spectrum} shows the dynamic spectrum.
Figure \ref{f:Y12の速度変化} (d) shows the time evolution of the blueshifted component with the maximum velocity of  254 km s$^{-1}$ and the duration of 14 minutes. Figure \ref{f:Y12の速度変化} (d) shows the blueshifted component accelerates and reaches its peak approximately 4 minutes after its appearance and then decelerates at a roughly constant rate, with a deceleration that is $\sim$0.5 times the surface gravity. This flare lacks a significant white-light flare emission, classified as a non-white-light flare (Figure \ref{f:Y12の速度変化}). The corresponding H$\alpha$ flare shows a complex light curve with several peaks and a long duration of 136 minutes, 
probably meaning a superposition of several flares. The blueshift event appears in the final phase of this series of flares with a potential H$\alpha$ peak. There is a corresponding white-light spike with an amplitude of approximately 0.01 associated with blue asymmetry in TESS light curve (Figure \ref{f:Y12の速度変化}),  but it was not judged as a flare in our flare detection algorithm because it does not have more than three consecutive points. During the series of the flares, we can see possible red asymmetry, e.g., in 80--100 min, as in Figure \ref{f:Y12_dynamic_spectrum}, but it was not significant based on the BIC threshold.

By following the same discussion as in Section \ref{subsubsec:Y8の起源}, the relatively high velocity of the blue asymmetry simply suggests that this blueshifted component is likely to indicate a prominence eruption. The almost constant deceleration, approximately 0.5 times the surface gravity, may suggest either the free fall of a prominence eruption that occurred at the limb and erupted at an angle of 60 degrees from the line of sight, or a prominence eruption that occurred at a height of 1.4 R$_{\text{star}}$ above the disc. The interesting point is that this blue asymmetry is seen in the latter half of a series of H$\alpha$ flares, following a possible red asymmetry. As shown in Figures \ref{f:Y12_dynamic_spectrum} (a) and \ref{f:Y12の速度変化} (b), this blue asymmetry appears approximately 40 minutes after the H$\alpha$ flare peak at $t\sim$75 min. This delayed blue asymmetry is similar to the event reported by \cite{Inoue2024}. \cite{Inoue2024} reported a blue asymmetry that occurred approximately 1 hour after the H$\alpha$ flare peak, and this delay time is comparable to this event.
Blue asymmetries of stellar flares are reported to begin to appear around at the peak of flares \citep[e.g.,][]{Vida2016,Honda2018,Maehara2021,Notsu2024}, so this kind of delayed case is relatively rare.
\cite{Inoue2024} proposed that the delay in this blue asymmetry could be due to (i) another flare/prominence eruption occurring in the decay phase (e.g., \citealt{Zirin1969}: \citealt{Mason2021} for solar observations) or (ii) a prominence erupted during the decay phase of the flare (e.g., \citealt{Kurokawa1987} for solar observations). 
Figure \ref{f:Y12の速度変化} (b) shows the increase in the equivalent width of the H$\alpha$ line center component at the onset of the blue asymmetry can be observed at $t\sim$115 min. This suggests that this blue asymmetry may be associated with a prominence eruption triggered by another flare in the latter half of the duration of flare Y12. Therefore, we cannot distinguish these two possibilities for this event as well.

In addition to these interpretations, the possible red asymmetry ($\Delta$BIC $\sim$ 0) that was below the BIC threshold ($\Delta$BIC $>$ 2) before the blueshift is not accompanied by significant white-light flare. This may suggest the occurrence of backward eruptions around the stellar limb (see Section \ref{subsec:Y6の起源}). The later blueshift could indicate falling material. On the other hand, the velocity during the blue asymmetry accelerates initially and then decelerates at a roughly constant rate. These velocity changes cannot be explained by backward eruptions because the blueshifted velocity should increase as the material falls.

\subsubsection{Flare Y20: Long-Duration Blue Asymmetry}
\label{subsec:Y20の起源}
\begin{figure}
        \centering
        \includegraphics[width=1\linewidth]{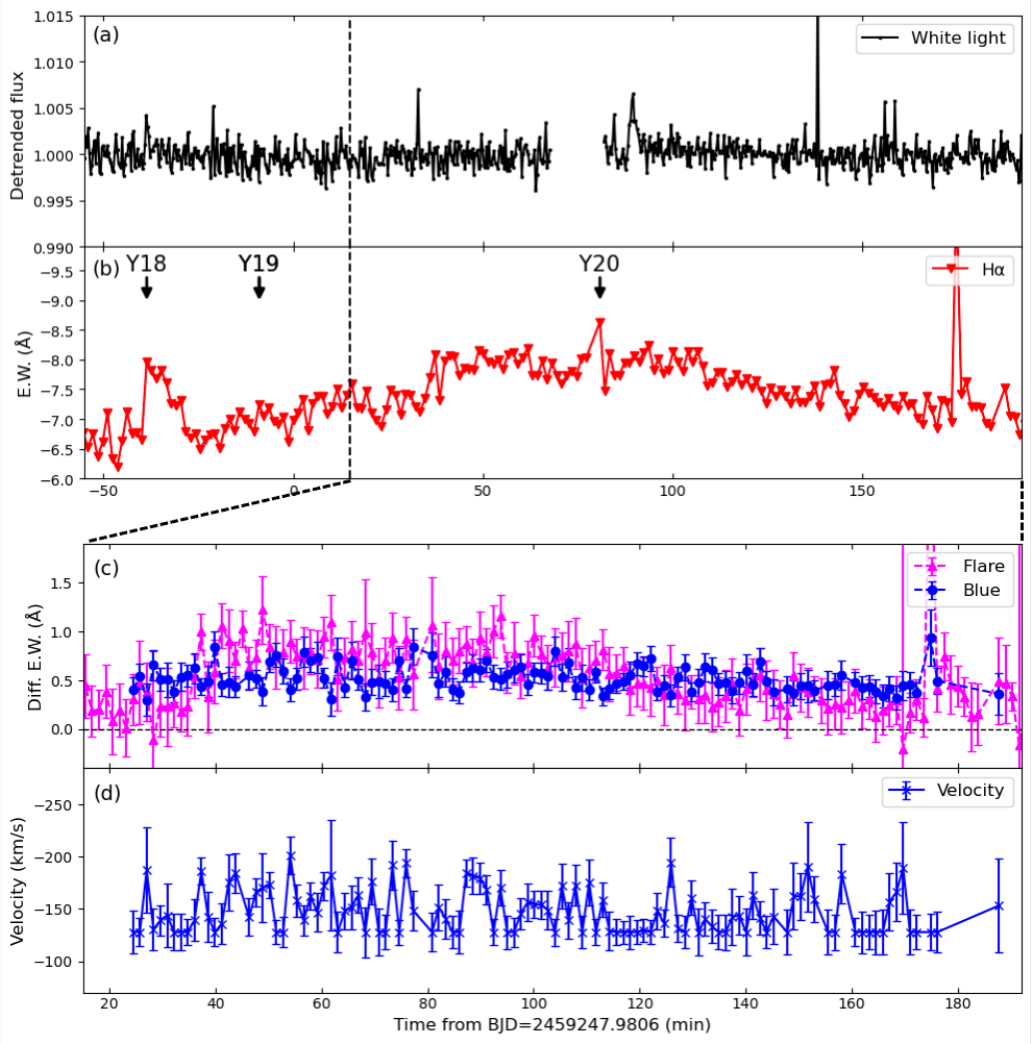}
        \caption{The same as Figure 11, but for the event of Y20. In (b), black arrows mark the peak times of flares Y18, Y19, and Y20, respectively}

        \label{f:Y20の速度変化}
    \end{figure}

In regard to the blue asymmetry during the flare Y20 on February 2, Figure \ref{f:Y20_dynamic_spectrum} shows the dynamic spectrum.
Figure \ref{f:Y20の速度変化} (b) shows that before Y20, there are consecutive flares, the impulsive H$\alpha$ flare Y18 and the gradual H$\alpha$ flare Y19. 
Also, flare Y20 is a non-white-light flare.
Almost throughout the H$\alpha$ flare, blue asymmetry is observed with the maximum velocity of 202±18 km s$^{-1}$.
We found the duration of asymmetry is 160 minutes, which is the longest among H$\alpha$ blueshift events reported in publications (the longest one was 120 min in \citealt{Honda2018}).
It is noted that the velocity was almost constant $\sim$150 km s$^{-1}$ for three hours.
These characteristics are similar to the blue asymmetry event in M-dwarf EV Lac reported by \cite{Honda2018}.

As discussed in \ref{subsubsec:Y8の起源}, the typical velocities of chromospheric upflows observed in solar flares (a few 10 km s$^{-1}$; \citealt{Svestka1962,Tei2018,Li2019,Huang2019}) are much smaller than the maximum velocity of the observed blueshifted components (202±18 km s$^{-1}$). This suggests that chromospheric upflows are unlikely to be the cause.

Next, we discuss the possibility of prominence eruptions. The duration of the asymmetry (160 minutes) is significantly longer than those of typical prominence/filament eruptions on the Sun (several minutes to tens of minutes; e.g., \citealt{Otsu2022,Otsu2024}) and those expected for M-dwarfs ($\sim$a few to tens of minutes).
Also, the blueshifted components maintain almost constant velocity for three hours (Figure \ref{f:Y20の速度変化} (d)). 
If this were to be a prominence eruption, there should be some velocity changes due to the gravity. 
These characteristics are inconsistent with prominence eruptions.

Then, could post-flare loops be possible? In the case of the Sun, post-flare loops sometimes last more than hours \citep[e.g.,][]{Liu2013,Song2016,Otsu2024}. This is consistent with the observed long-duration asymmetry with duration of 160 minutes. Additionally, the H$\alpha$ spectrum exhibits a blue asymmetry profile with a missing red wing (or possible redshifted absorption), suggesting the possibility of absorption by post-flare loops, similar to the red wing absorption seen in solar post-flare loops (e.g., \citealt{Otsu2022,Otsu2024}). In addition, the H$\alpha$ light curve shows that gradual flares Y19 and Y20 occur following Y18. This may resemble the ``peak-bump" structure observed in the TESS white-light light curve and Ca II H \& K lines \citep{Kowalski2013,Howard&MacGregor2022}. These post-flare bump features are sometimes interpreted as emissions from post-flare loops
\citep{Heinzel&Shibata2018,Kai2023}. The Sun-as-a-star analysis of post-flare loops in the H$\alpha$ line suggests that post-flare secondary peak would also be observed in the H$\alpha$ line \citep{Otsu2024}.
Based on this interpretation, the Y19--20 events might represent post-flare loops emerging after the Y18 flare. However, the peak-bump structure and redshift absorption each assumes different mechanisms--emission from post-flare loops and absorption by post-flare loops, respectively--resulting in a contradiction between these interpretations.
Also, Figure \ref{f:Y20_dynamic_spectrum} (c) shows that the line profile exhibits a bluewing enhancement with extended velocity (or expanded ``bump-like" shape centered at approximately $-$250 km s$^{-1}$). 
This shape cannot be solely explained by redshift absorption from the post-flare loop because the red wing absorption itself cannot make the bump-like feature in the blue wing. 
Therefore, all of these three hypothesis introduced in Section \ref{subsubsec:解釈レビュー} cannot simply explain the observed properties. To fully explain the observation, we may need to consider an alternative interpretation and/or some situation specific to M-dwarfs or this event.

\subsection{Origin of Rapid, Short-Duration Red Asymmetry}
\label{subsec:Y6の起源}
Flare Y6 on 2021 January 25 shows the rapid, short duration red asymmetry. Although red asymmetry is not the main focus of this paper, we discuss its origin in detail here because of its short-duration and the time variation resembling a backward prominence eruption.

\begin{figure}
        \centering
        \includegraphics[width=1\linewidth]{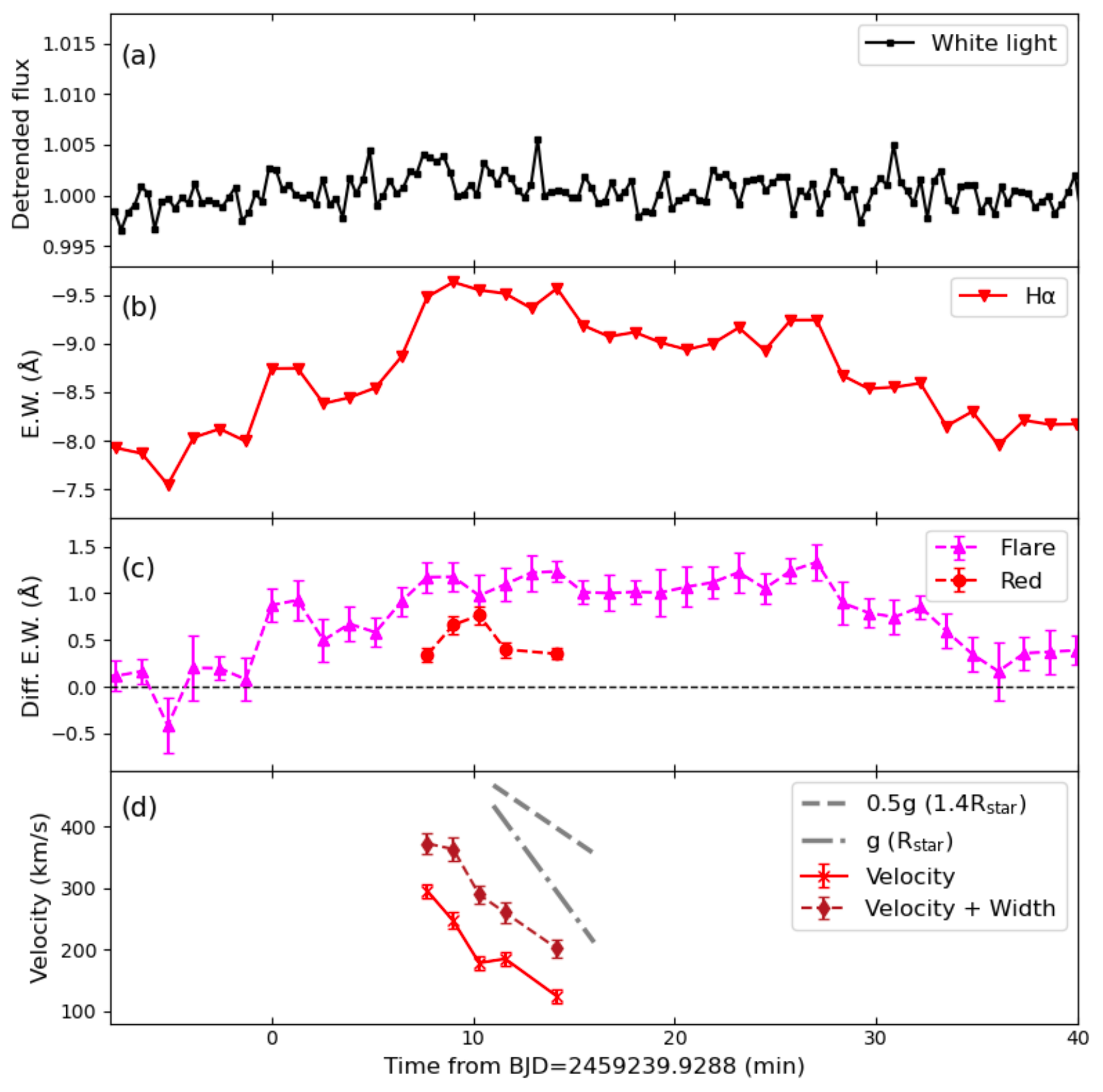}
        \caption{The same as Figure 12, but for the event of Y6. In this case, the focus is on the redshifted components.}
        
        \label{f:Y6の速度変化}
\end{figure}

Figure \ref{f:Y6_dynamic_spectrum} shows the dynamic spectrum of the Y6 event and Figure \ref{f:Y6の速度変化} shows the time evolution of the equivalent width and velocity of the redshifted components.  
Figure \ref{f:Y6の速度変化} (a) shows that a small but significant increase in white light can be observed at the peak of the H$\alpha$ light curve in Y6. 
The bolometric flare energy $E_{\text{flare}}$ is 5.80×10$^{30}$ erg, while the H$\alpha$ flare energy $E_{\text{H}\alpha}$ is 2.3×10$^{30}$ erg.
It is noted that, the ratio of H$\alpha$ energy to bolometric energy is $\sim$0.5, which is relatively higher than the typical value of 0.01 to 0.1 \citep[see Figure 18; e.g.,][]{Namekata2024}.  
The maximum velocity of these redshifted components is 295±12 km s$^{-1}$ (Table \ref{tab:Flares showing red/blue asymmetry}).
Many of previous studies have shown that the typical redshift velocity in M-dwarfs is 100-200 km s$^{-1}$ \citep{Vida2016,Wu2022,Wollmann2023,Notsu2024}, while \cite{Namizaki2023} reported a maximum velocity of $\sim$500 km s$^{-1}$.
The redshift velocity of this event is relatively faster than--but consistent with--those in these previous studies. 
We found that the duration of this red asymmetry is 6 minutes, which is the shortest among H$\alpha$ redshift events reported in publications (see Section \ref{subsec:高速短時間非対称イベント}).

The primary origins of red asymmetry in M-dwarfs are considered as chromospheric condensation and/or the downward flow of post-flare loops (e.g., \citealt{Namizaki2023}), but sometimes referred to backward prominence eruptions \citep[e.g.,][]{Wu2022}.
In the case of the Sun, post-flare loops occur in the decay phase of a flare, and their duration often exceeds several tens of minutes. 
On the other hand, red asymmetry in Y6 is associated with white-light flare, an impulsive phase, at the peak of the H$\alpha$. 
This suggests that post-flare loops are unlikely to be the cause. 
Next, 
chromospheric condensation is thought to be formed by the compression of the chromosphere due to the high injection of high-energy non-thermal electrons \citep{Allred2006, Longcope2014, Kowalski2017a, Kowalski2022b}. Therefore, a strong correlation with white-light flares originating from the chromospheric condensation region is expected \citep{Namizaki2023}. However, despite the velocity reaching 295±12 km s$^{-1}$, the accompanying white-light flare in this event is extremely faint (barely detectable). This suggests that chromospheric condensation is qualitatively unlikely to be the cause, although the possibility remains.

On the other hand, the typical velocity of solar prominence eruptions ranges from 10 to 400 km s$^{-1}$ \citep{Gopalswamy2003,Seki2021}, and some have durations of around 5 minutes (e.g., \citealt{Otsu2022}). The maximum velocity of 295±12 km s$^{-1}$ and duration of approximately 6 minutes in this event are relatively fast and short but within the typical velocity and duration of solar prominence eruptions. 
In addition, $E_{\text{flare}}$ (5.8×10$^{30}$ erg) and $E_{\text{H}\alpha}$ (2.3×10$^{30}$ erg) differ by only a factor of 2, which is strange compared to typical stellar flares.
Since the TESS band includes the H$\alpha$ line, the increase in white light can be almost entirely explained by the increase in H$\alpha$ emission.  
This may suggest that the flare occurs around or behind the stellar limb and prominence eruption is the primary source of the H$\alpha$ radiation, which is consistent with the scenario of a backward prominence eruption.

Under the assumption that the observed redshift is a backward prominence eruption, we speculate some dynamics of the stellar eruption.
Figure \ref{f:Y6の速度変化} (d) shows that the component reaches its maximum velocity already, without any acceleration phase observed. 
This suggests that the plasma acceleration is completed in less than 1 minute. Additionally, it decelerates over approximately 5 minutes, at a rate roughly equal to 0.5 times the surface gravitational acceleration. This may suggest the free fall of a prominence eruption that occurred at the limb and erupted at an angle of 120 degrees from the line of sight. 

\subsection{Implications to Stellar CMEs}\label{sec:Implications to Stellar CMEs}

In this section, we focus on the detected possible prominence eruptions and discuss their potential development into CMEs. As discussed in Section \ref{subsec:OriginsOfBlueAsymmetry} and Section \ref{subsec:Y6の起源}, the red/blue asymmetries observed during the three flares, Y6, Y8, and Y12, suggest prominence eruptions. The maximum velocities of these eruptions are 295±12 km s$^{-1}$, 445±36 km s$^{-1}$, and 254±43 km s$^{-1}$, respectively. Since these velocities are below the surface escape velocity of YZ CMi ($\sim$600 km s$^{-1}$), it is unclear whether the majority of prominence themselves developed into CMEs. However, in the case of the Sun, the average velocity of CMEs (610 km s$^{-1}$) is approximately 8 times the velocity of the associated prominence eruptions, approximately 80 km s$^{-1}$ \citep{Gopalswamy2003}. Assuming a similar relationship between prominence eruptions and CMEs on YZ CMi, the upper layers of hot plasma associated with the prominence eruptions in Y6 \& Y8, and Y12 could be accelerated to velocities of approximately 1800--3500 km/s. These velocities are much higher than the surface escape velocity of YZ CMi ($\sim$600 km s$^{-1}$). This suggests that these prominence eruptions are likely to have led to CMEs.

In addition to the central Gaussian velocity, we here mention the velocity dispersion for each event Y6, Y8, and Y20. As we described above, the observed maximum velocities of these events are below the escape velocity of YZ CMi. However, these blueshifted components are broadened more than velocity resolution $\sim$70 km s$^{-1}$ \citep{Matsubayashi2019}. In Figure \ref{f:Y8の速度変化} and Figure \ref{f:Y12の速度変化} (d), the navy diamonds indicate the evolution of the central velocity plus the Gaussian line width $\sigma$. Similarly, in Figure \ref{f:Y6の速度変化} (d), the brown diamonds show the same evolution. This can be interpreted as the velocity distribution of the expanding prominence \citep{Namekata2024}. Based on this interpretation, we can assume that the width plus central velocity can be regarded as the maximum velocity component of the eruptive prominence. In particular, we found that the velocity of Y8 almost reaches the escape velocity within the error bar, while Y6 and Y12 do not show a similar trend to Y8. This can suggest that some fast components of the eruptive prominence themselves could have developed into CMEs.

The velocity changes of the possible prominence eruptions tell us its evolution. As for the Y8, the velocity reaches its maximum at the end of the duration, with no observed deceleration afterward. 
This suggests that the accelerated prominence expanded spatially and became optically thin.
On the other hand, although there is no velocity component faster than $v_{\text{esc}}$, Y6 and Y12 show no subsequent blue/red asymmetry after these red/blue asymmetries. This suggests that the prominence eruptions in Y6 and Y12 did not fall back to the stellar surface. 

Next, we discuss the proportion of flares accompanied by prominence eruptions. In this research, 3 out of 27 H$\alpha$ flares ($\sim$10\%) suggest prominence eruptions. This ratio is comparable to the value ($\sim$15\%) reported by \cite{Notsu2024} for M-dwarfs. However, the association rate of prominence eruptions in these M-dwarfs is significantly lower compared to that in a G-dwarf ($\sim$60\%; \citealt{Namekata2024}). This difference could be caused by several factors. First, there is a difference in the observable energy range of flares. G-dwarfs have brighter surfaces compared to M-dwarfs, making it harder to detect lower-energy flares in M-dwarfs. In the case of the Sun, higher-energy flares are more likely to be associated with eruptive events and have higher velocities \citep{Yashiro&Gopalswamy2009}. The flares detected in G-dwarfs are relatively high-energy, potentially resulting in a higher apparent  association rate of prominence eruptions compared to M-dwarfs. Second, there is a difference in the brightness contrast between the filament on the disk and the stellar surface in M-dwarfs. The surface of M-dwarfs is darker, reducing the contrast with filament on the disk, making them harder to observe \citep{Leitzinger2022}. This may have lowered the detection rate of filaments. Third, there is a difference in the coronal magnetic field strength and topology. \cite{Alvarado-Gómez2018} suggested through 3D magnetohydrodynamic simulations that CMEs could be suppressed in regions with strong overlying magnetic fields. Fast-rotating early-mid M-dwarfs like YZ CMi have dipole-dominated or multipolar large-scale fields \citep{Gastine2013}. 
Zeeman Doppler Imaging observations by \cite{Morin2008, Lang2014} reported that YZ CMi had a large-scale, axisymmetric poloidal magnetic field in 2007 and 2008, with a much stronger magnetic flux than that of  G-dwarfs.
Therefore, the suppression of prominence eruptions due to a strong overlying magnetic field might occur more frequently in M-dwarfs.

Finally, we consider the frequency of prominence eruptions/CMEs. The total H$\alpha$ observation period  was 2.8 days. During this time, we interpreted three events (Y6, Y8, Y12) as prominence eruptions. The occurrence frequency of prominence eruptions in this study is $\sim$1.1 events per day, which is higher than the frequency of prominence eruptions on the Sun (0.1 to 0.4 events per day; \citealt{Gopalswamy2015}) and G-dwarfs ($\sim0.3$ events per day; \citealt{Namekata2024}). As we discussed in Sections \ref{subsubsec:Y8の起源} and \ref{subsec:Y6の起源}, the asymmetries in flares Y6 and Y8 which lasted less than 10 minutes, are likely caused by rapid, short-duration prominence eruptions. Since 2 out of the 3 events that we interpreted as prominence eruptions are short-duration, such events may not be rare. This suggests that previous studies of M-dwarf blue asymmetries from low time cadence spectroscopy may have underestimated the frequency of prominence eruptions/CMEs due to insufficient temporal resolution. On the other hand, the proportion of flares accompanied by prominence eruptions is comparable to the proportion reported by \cite{Notsu2024}. We should note that the wavelength resolution (R$\sim$2000) of our observation is lower than that of \cite{Notsu2024} (R$\sim$32000). Therefore, we may have missed low-velocity prominence eruptions that \cite{Notsu2024} could detect. For future work, high time cadence ($\sim$1 min) and high dispersion spectroscopy are necessary to more accurately estimate the occurrence frequency of prominence eruptions/CMEs.

\section{Summary and Conclusions}\label{sec:Summary and Conclusions}

In this study, we conducted simultaneous photometric and spectroscopic observations of the active M-dwarf YZ CMi. As a result, we detected 27 H$\alpha$ flares and 130 white-light flares, while 17 H$\alpha$ flares are not associated with white-light flares. The H$\alpha$ flare energy $E_{\rm H \alpha}$ ranges from 1.7 $\times$ 10$^{29}$ to 3.8 $\times$ 10$^{32}$ erg, and the duration ranges from 8 to 319 minutes. The bolometric energy $E_{\text{flare}}$ ranges from 4.7 $\times$ 10$^{30}$ to 1.2 $\times$ 10$^{34}$ erg, and the duration ranges from 1.7 to 235.3 minutes. We introduced a new criteria based on the Bayesian Information Criterion (BIC) to identify asymmetries of H$\alpha$ line profile. Among them, 5 flares (Y4, Y5, Y6, Y16, Y21) show red asymmetries, and 3 flares (Y8, Y12, Y20) show blue asymmetries in the H$\alpha$ line profile. For red asymmetry, the maximum velocity of the redshifted components ($v_{\text{asym,max}}$) ranges from 188 to 400 km s$^{-1}$, with durations ($\tau_{\text{asym}}$) ranging from 6 to 319 minutes. Similarly, for blue asymmetry, the maximum velocity of the blueshifted components ($v_{\text{asym,max}}$) also ranges from 188 to 400 km s$^{-1}$, with durations ($\tau_{\text{asym}}$) ranging from 6 to 319 minutes. The physical interpretations of the red/blue asymmetries in four events (Y6, Y8, Y12, Y20) are in particular discussed in Section \ref{subsec:OriginsOfBlueAsymmetry}-\ref{subsec:Y6の起源}. The blue asymmetries are discussed in terms of chromospheric upflow, prominence eruption, and apparent blue asymmetries from the post-flare loop absorption, while red asymmetries are in terms of chromospheric condensation, post-flare loop, and backward prominence eruption. Overall, the observed relatively high blueshifted velocities (200--450 km s$^{-1}$) prefer the possibility of prominence eruptions for almost all cases.
On the other hand, except for the observed velocity, we also found large diversities in the timing and duration of blue asymmetries and their association with the white-light flares. In particular, we discovered rapid, short-duration blue (Y8) and red (Y6) asymmetries with durations of $\sim$5 minutes, thanks to an unprecedented high time-cadence($\sim$1 minute) spectroscopy resolving H$\alpha$ profile.
Both durations of the asymmetries are the shortest among H$\alpha$ observations reported for M-dwarfs in publications. The obtained diversities enable us to constrain their origins. We suggest that blue asymmetries from Y8 and Y12 are likely originated from prominence eruptions. Also, the red asymmetry from Y6 is likely a backward prominence eruption occurring at the limb. Y20 is the long-duration blue asymmetry lasting 160 minutes. Several aspects indicate a possibility of a post-flare loop, but there are still inconsistencies, and further study is needed. For future work, modeling studies based on observational data incorporating radiative transfer (e.g., \citealt{Leitzinger2022}) and further the Sun-as-a-star analysis are necessary for more detailed discussions.

The occurrence frequency of prominence eruptions is $\sim$1.1 events per day. Our discovery of rapid, short-duration prominence eruptions (Y8 and Y6) indicate that previous studies with a low time cadence could have missed those events and as a result underestimate the occurrence frequency of the prominence eruptions/CMEs.  
Also, the low time cadence observations would miss the fast components of the prominence, which results in the underestimation of the velocity and kinetic energy of prominence eruptions (and CMEs). This relatively high velocity observed in this study may be able to fill a gap in scaling relations for flare energy and kinetic energy between the Sun and stars \citep[e.g.,][]{Moschou2019}.

Also, in paper II, we will be focusing on the statistical viewpoints of these asymmetric events. We will investigate duration, energy, the association of white-light flares, and the rotational phase for the eruptive/non-eruptive events and discuss further what could be the source of the asymmetry.

\section*{Acknowledgment}
We thank K. Shibata, T. Otsu, and A. Asai for their valuable comments and discussions.
This research was supported by JSPS (Japan Society for the Promotion of Science) KAKENHI Grant Numbers  21J00316 (K.N.), 20K04032, 24K00685 (H.M.), and 24H00248 (Y.K., K.N., H.M., and D.N). Y.N. acknowledges the funding support from NASA ADAP 80NSSC21K0632, and NASA TESS Cycle 6 80NSSC24K0493.
The spectroscopic data used in this paper were obtained through the program 21A-N-CN03 (PI: K.N.) with the 3.8m Seimei telescope, which is located at Okayama Observatory of Kyoto University.
This paper includes data collected with the TESS mission, obtained from the MAST data archive at the Space Telescope Science Institute (STScI). Funding for the TESS mission is provided by the NASA Explorer Program.
STScI is operated by the Association of Universities for Research in Astronomy, Inc., under NASA contract NAS 5-26555. 
Some of the data presented in this paper were obtained from the Mikulski Archive for Space Telescopes (MAST) at the Space Telescope Science Institute. The specific observations analyzed can be accessed via\dataset[10.17909/escv-3665]{https://doi.org/10.17909/escv-3665}. The authors acknowledge ideas from the participants in the workshop ``Blazing Paths to Observing Stellar and Exoplanet Particle Environments" organized by the W.M. Keck Institute for Space Studies.

\facilities{Seimei telescope, Transiting Exoplanet Survey Satellite (TESS)}

\software{\textsf{astropy} \citep{Astropy2018} , \textsf{IRAF} \citep{IRAF1986}, \textsf{PyRAF} \citep{Pyraf2012}}

\clearpage

\appendix

\section{Appendix: Velocity Determination Accuracy}\label{Appendix:Velocity Determination Accuracy}

We fitted the normalized spectrum in the quiescent state on the day of the events showing asymmetry, using a Voigt function. From this fit, we calculated the daily mean ($\lambda_{\text{Voigt, H}\alpha}$) and the standard deviation ($\sigma_{\text{v}}$) of the fitted function's line center. Table \ref{tab:Flares showing red/blue asymmetry} lists the values of $\lambda_{\text{Voigt, H}\alpha}$ and $\sigma_{\text{v}}$ for each day. The table shows that $\sigma_{\text{v}}$ is approximately 1-2 km s$^{-1}$. This value almost corresponds to the variation in the observed velocity solution during observations but is much smaller than the Gaussian fitting error used to estimate the velocity error of the asymmetric components (10-50 km s$^{-1}$; Table \ref{tab:Flares showing red/blue asymmetry}). Although $\lambda_{\text{Voigt, H}\alpha}$ varies within a range of 0.1 {\AA} ($\sim$ 4.5 km s$^{-1}$), this value is comparable to the rotational velocity of YZ CMi ($v\sin i \sim$ 4.5 km s$^{-1}$; \citealt{Reiners2007}).


\begin{table}[ht]
\centering
{%
\begin{threeparttable}
\caption{Voigt Fit H$\alpha$ Line Center Measurements}
\label{tab:voigt_fit_Halpha_line_center}
\begin{tabular}{ccc}
\hline\hline
Date & $\lambda_{\text{Voigt, H}\alpha}$ (Å) & $\sigma_{\text{v}}$ (km s$^{-1}$) \\
\hline
2021 Jan 24 & 6562.81 & 1.82 \\
2021 Jan 25 & 6562.74 & 1.44 \\
2021 Jan 28 & 6562.72 & 1.26 \\
2021 Jan 30 & 6562.69 & 0.99 \\
2021 Jan 31 & 6562.74 & 1.16 \\
2021 Feb 02 & 6562.74 & 1.48 \\
2021 Feb 03 & 6562.87 & 1.13 \\
\hline
\end{tabular}
\begin{tablenotes}
\item[a] The mean center wavelength (Å) of the H$\alpha$ line determined by the Voigt fit in the quiescent state.
\item[b] $\sigma$ is the standard deviation of the fit center in the quiescent state.

\end{tablenotes}
\end{threeparttable}%
}
\end{table}

\section{Appendix: Flare Light Curves and Dynamic Spectrum in the H$\alpha$ Line for Which Asymmetries Were Not Detected}\label{sec:Appendix}
\begin{figure}
    \centering
    \includegraphics[width=0.8\linewidth]{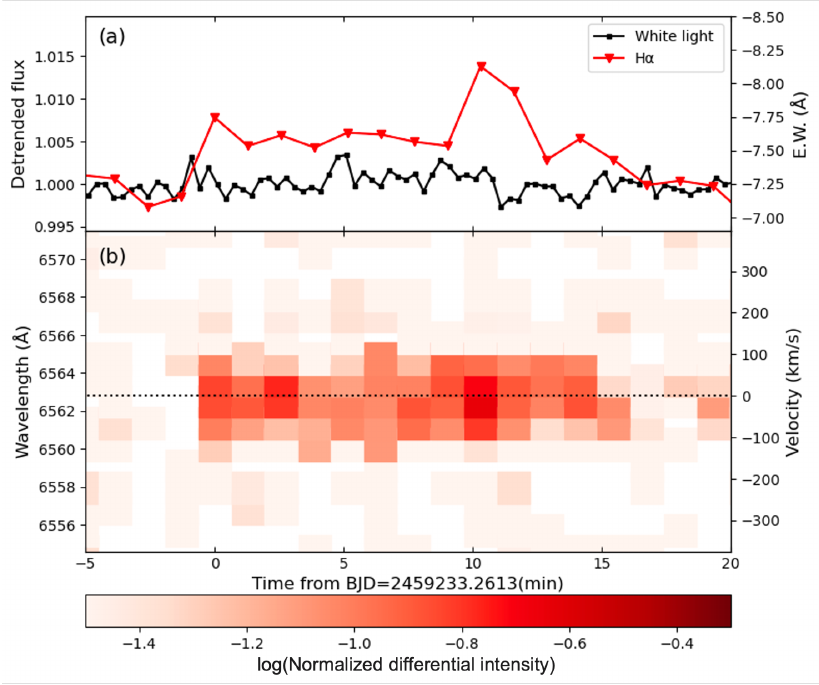}
    \caption{(a) Light curves of white light and H$\alpha$ line for flare Y1. The black solid line represents the detrended light curve of the white light emission showing long-term variations, and the red solid line represents the light curve of the H$\alpha$ line. The left vertical axis represents the detrended flux ($F/F_{\text{star}}$), and the right vertical axis represents the equivalent width of the H$\alpha$ line.(b) Temporal evolution of the H$\alpha$ line profile for Y1. The left vertical axis represents the wavelength, and the right vertical axis shows the Doppler velocity relative to the central wavelength of H$\alpha$ (6562.8\AA). The horizontal axis represents the elapsed time (min) from the start of the flare, with BJD=2459233.2613 as the reference point. The color bar indicates the logarithm of the normalized intensity in the differential spectrum. Black arrows mark the time of maximum $\Delta$BIC, and red bidirectional arrows indicate the duration of red asymmetry.
    (The complete figure set (15 images) is available in the online journal.)}
    \label{f:Y1_dynamic_spectrum}
\end{figure}

In this appendix, we show flare light curves and flare dynamic spectrum in H$\alpha$ line for which asymmetries were not detected (see Section \ref{subsec:Asymmetry_detection} and
Table \ref{tab:HalphaFlareList}). We do not discuss these events in detail, but the figures are shown in Figure \ref{f:Y1_dynamic_spectrum}. The figures for Y18 and Y19, which do not show asymmetries, are included in Figure 
\ref{f:Y20_dynamic_spectrum}.

\figsetstart
\figsetnum{16}  
\figsettitle{Light curves and dynamic spectrum in H$\alpha$ line for which asymmetries were not detected}  

\figsetgrpstart
\figsetgrpnum{16.1} 
\figsetgrptitle{Light curves and dynamic spectrum in H$\alpha$ line for Y1} 
\figsetplot{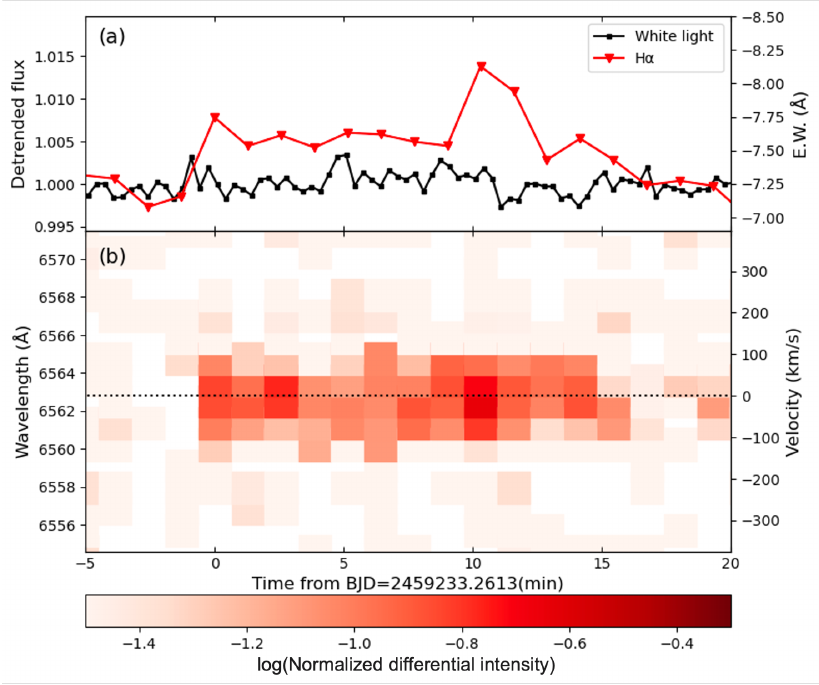}
\figsetgrpnote{(a) Light curves of white light and H$\alpha$ line for flare Y1. The black solid line represents the detrended light curve of the white light emission showing long-term variations, and the red solid line represents the light curve of the H$\alpha$ line. The left vertical axis represents the detrended flux ($F/F_{\text{star}}$), and the right vertical axis represents the equivalent width of the H$\alpha$ line.(b) Temporal evolution of the H$\alpha$ line profile for Y1. The left vertical axis represents the wavelength, and the right vertical axis shows the Doppler velocity relative to the central wavelength of H$\alpha$ (6562.8\AA). The horizontal axis represents the elapsed time (min) from the start of the flare, with BJD=2459233.2613 as the reference point. The color bar indicates the logarithm of the normalized intensity in the differential spectrum. Black arrows mark the time of maximum $\Delta$BIC, and red bidirectional arrows indicate the duration of red asymmetry.}  
\figsetgrpend

\figsetgrpstart
\figsetgrpnum{16.2} 
\figsetgrptitle{Light curves and dynamic spectrum in H$\alpha$ line for Y2} 
\figsetplot{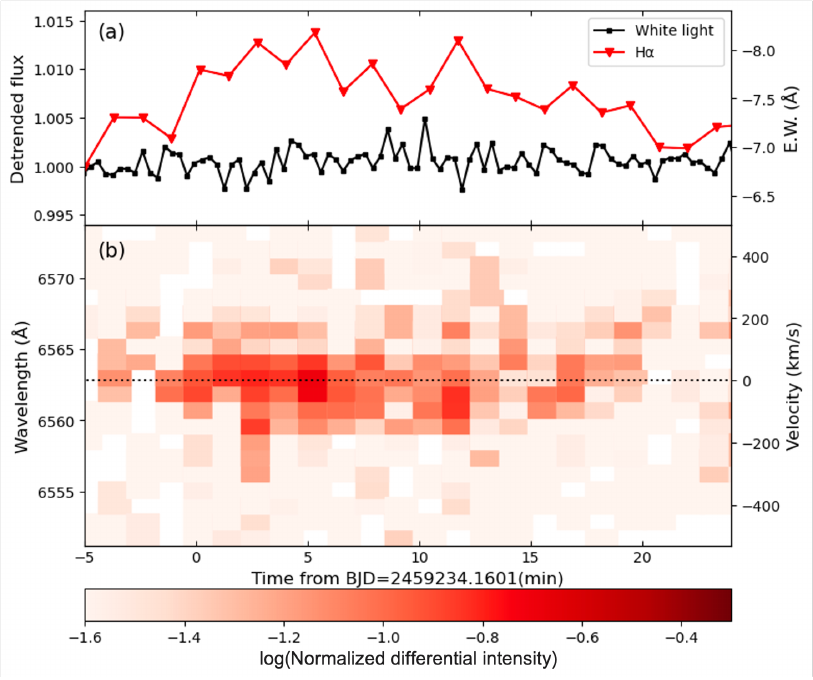}  
\figsetgrpnote{The same as Figure 16.1, but for the event of Y2.} 
\figsetgrpend

\figsetgrpstart
\figsetgrpnum{16.3} 
\figsetgrptitle{Light curves and dynamic spectrum in H$\alpha$ line for Y3} 
\figsetplot{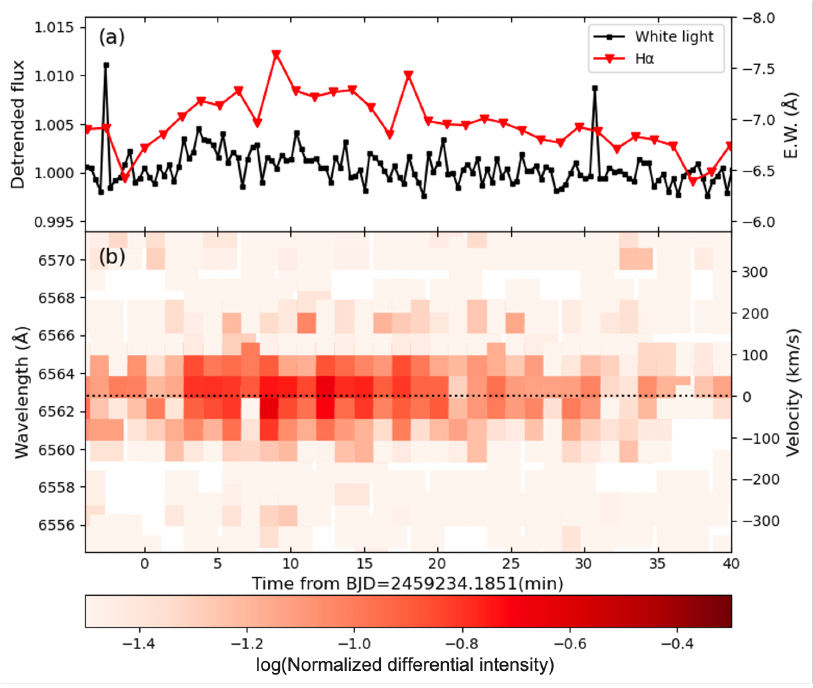}  
\figsetgrpnote{The same as Figure 16.1, but for the event of Y3.}  
\figsetgrpend

\figsetgrpstart
\figsetgrpnum{16.4} 
\figsetgrptitle{Light curves and dynamic spectrum in H$\alpha$ line for Y7} 
\figsetplot{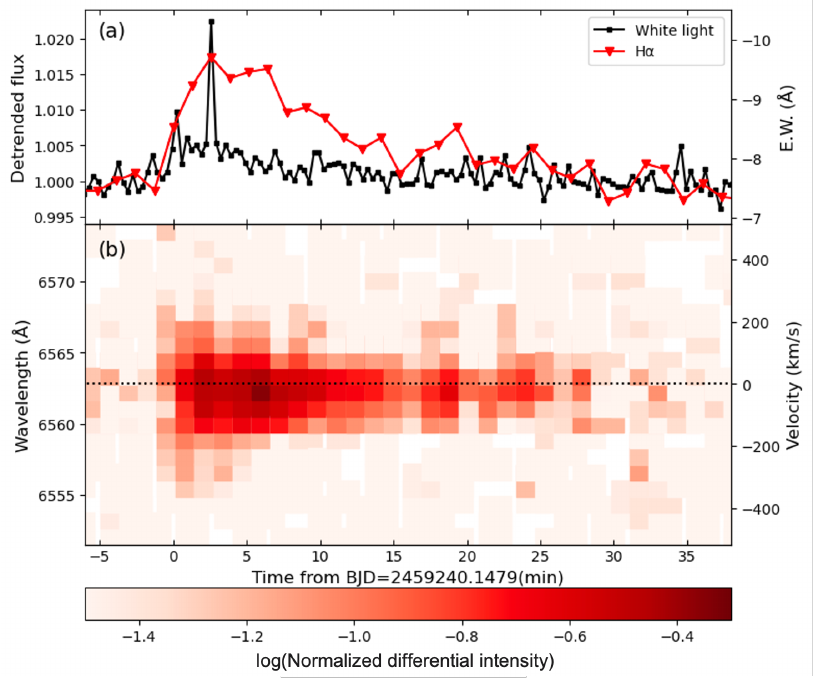}  
\figsetgrpnote{The same as Figure 16.1, but for the event of Y7.}  
\figsetgrpend

\figsetgrpstart
\figsetgrpnum{16.5} 
\figsetgrptitle{Light curves and dynamic spectrum in H$\alpha$ line for Y9} 
\figsetplot{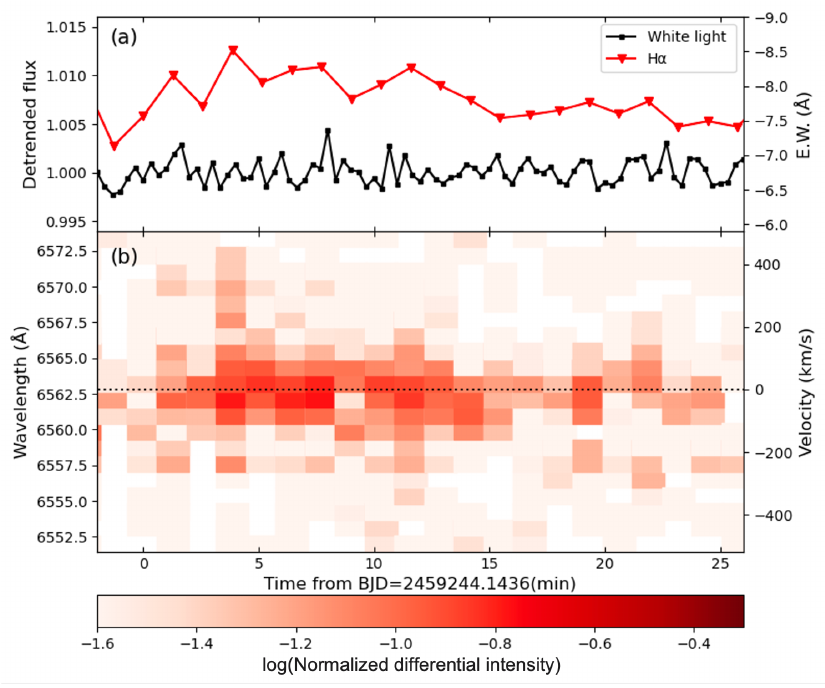}  
\figsetgrpnote{The same as Figure 16.1, but for the event of Y9.}  
\figsetgrpend

\figsetgrpstart
\figsetgrpnum{16.6} 
\figsetgrptitle{Light curves and dynamic spectrum in H$\alpha$ line for Y10} 
\figsetplot{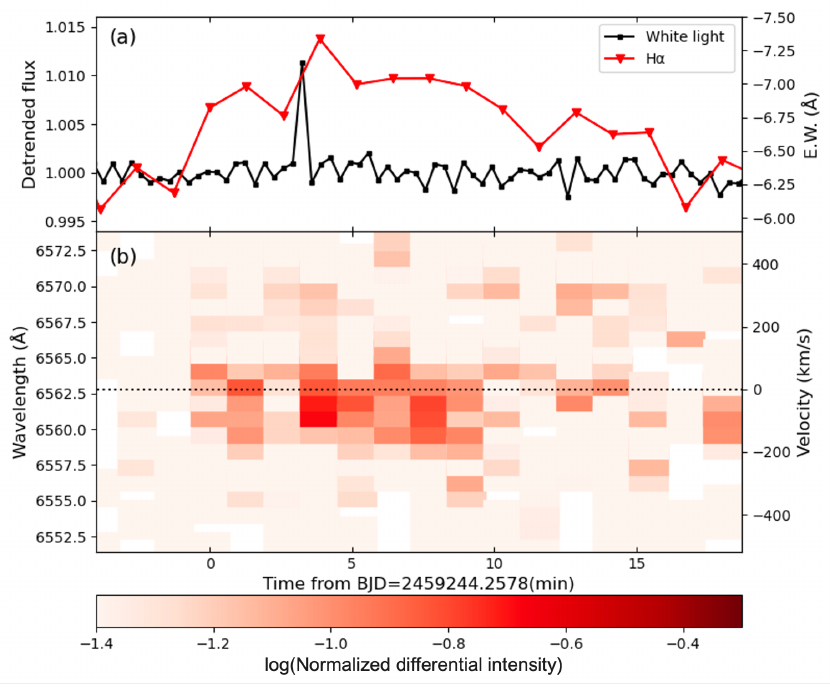}  
\figsetgrpnote{The same as Figure 16.1, but for the event of Y10.} 
\figsetgrpend

\figsetgrpstart
\figsetgrpnum{16.7} 
\figsetgrptitle{Light curves and dynamic spectrum in H$\alpha$ line for Y11} 
\figsetplot{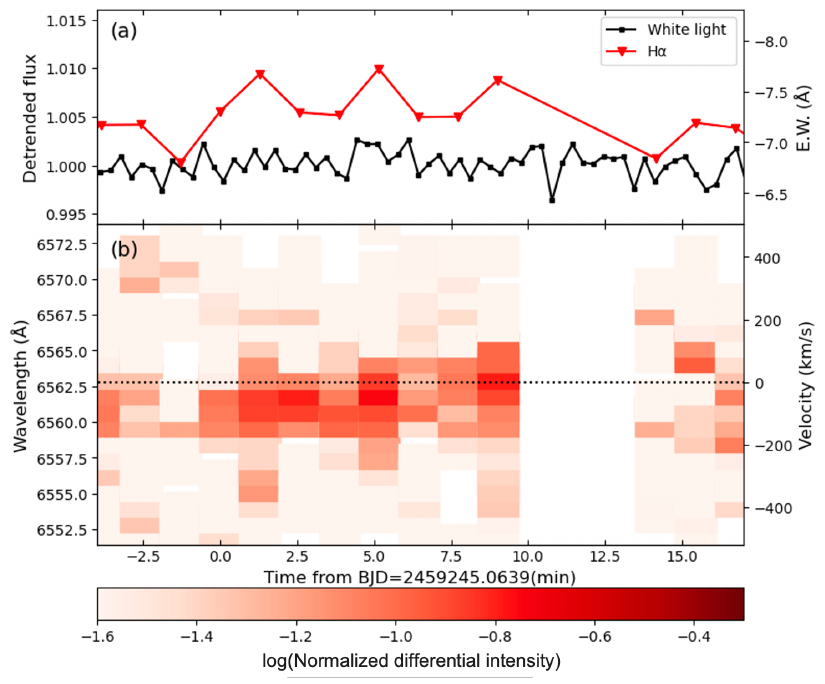}  
\figsetgrpnote{The same as Figure 16.1, but for the event of Y11.} 
\figsetgrpend

\figsetgrpstart
\figsetgrpnum{16.8} 
\figsetgrptitle{Light curves and dynamic spectrum in H$\alpha$ line for Y13 and 14} 
\figsetplot{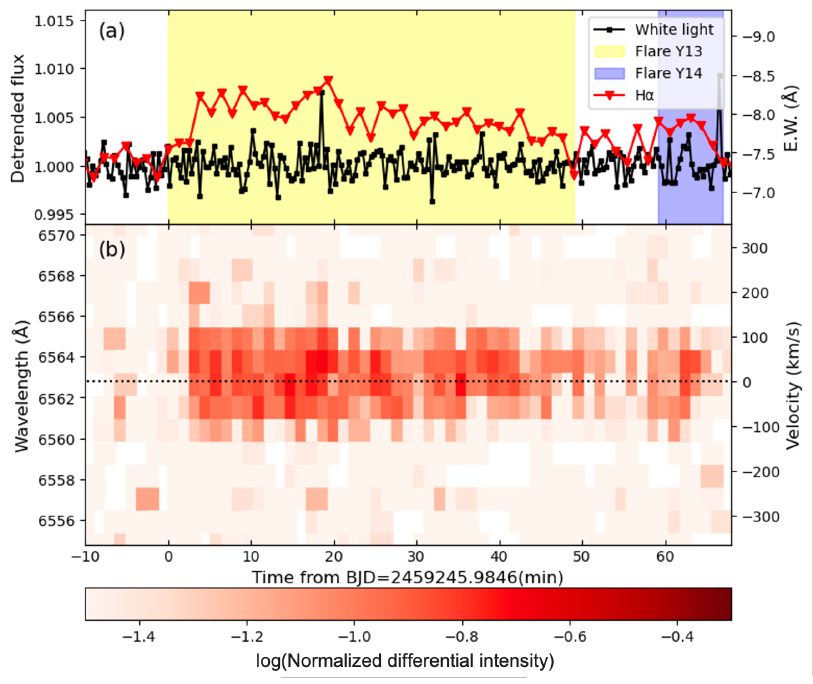}  
\figsetgrpnote{The same as Figure, but for the events of Y13 and Y14. The yellow shaded area shows the duration of Y13, and the blue shaded area shows the duration of Y14.}  
\figsetgrpend

\figsetgrpstart
\figsetgrpnum{16.9} 
\figsetgrptitle{Light curves and dynamic spectrum in H$\alpha$ line for Y15} 
\figsetplot{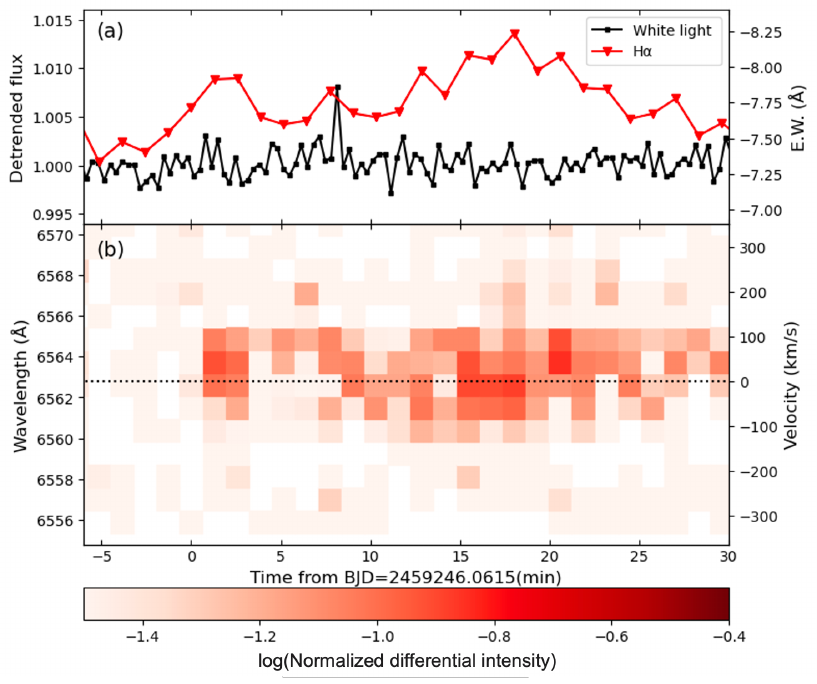}  
\figsetgrpnote{The same as Figure, but for the events of Y15.}  
\figsetgrpend

\figsetgrpstart
\figsetgrpnum{16.10} 
\figsetgrptitle{Light curves and dynamic spectrum in H$\alpha$ line for Y17} 
\figsetplot{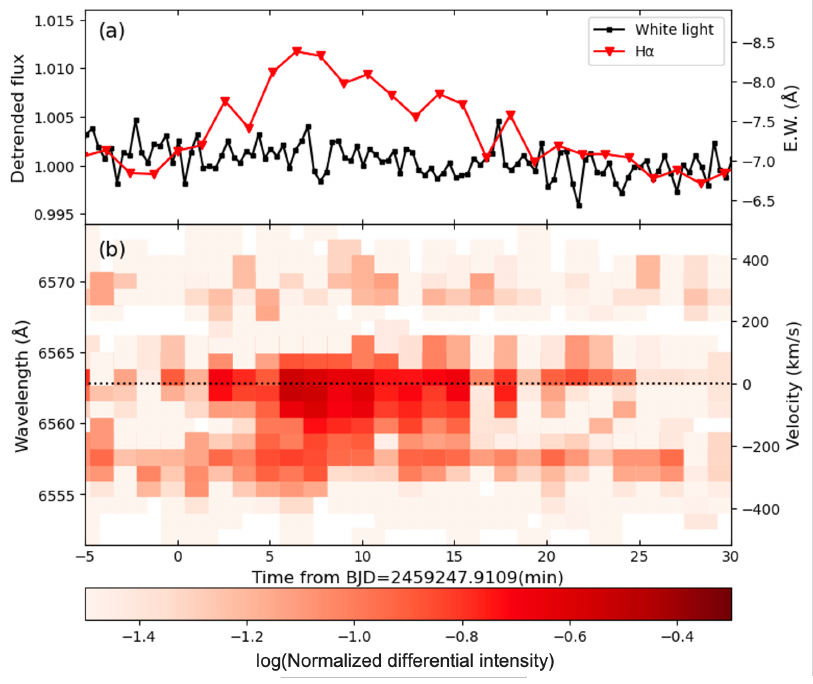}  
\figsetgrpnote{The same as Figure, but for the events of Y17.} 
\figsetgrpend

\figsetgrpstart
\figsetgrpnum{16.11} 
\figsetgrptitle{Light curves and dynamic spectrum in H$\alpha$ line for Y22} 
\figsetplot{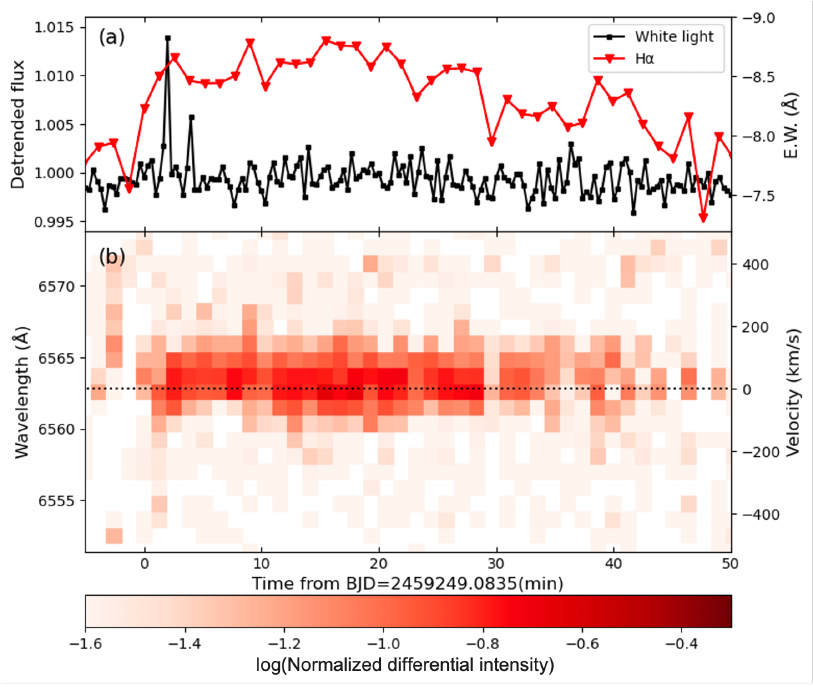}  
\figsetgrpnote{The same as Figure, but for the events of Y22.} 
\figsetgrpend

\figsetgrpstart
\figsetgrpnum{16.12} 
\figsetgrptitle{Light curves and dynamic spectrum in H$\alpha$ line for Y23} 
\figsetplot{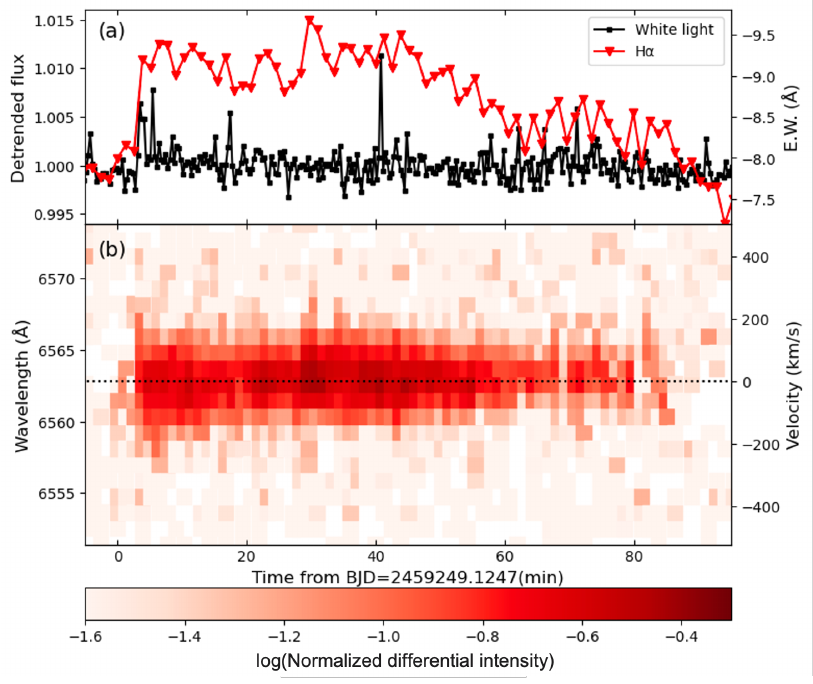}  
\figsetgrpnote{The same as Figure, but for the events of Y23.}  
\figsetgrpend

\figsetgrpstart
\figsetgrpnum{16.13} 
\figsetgrptitle{Light curves and dynamic spectrum in H$\alpha$ line for Y24} 
\figsetplot{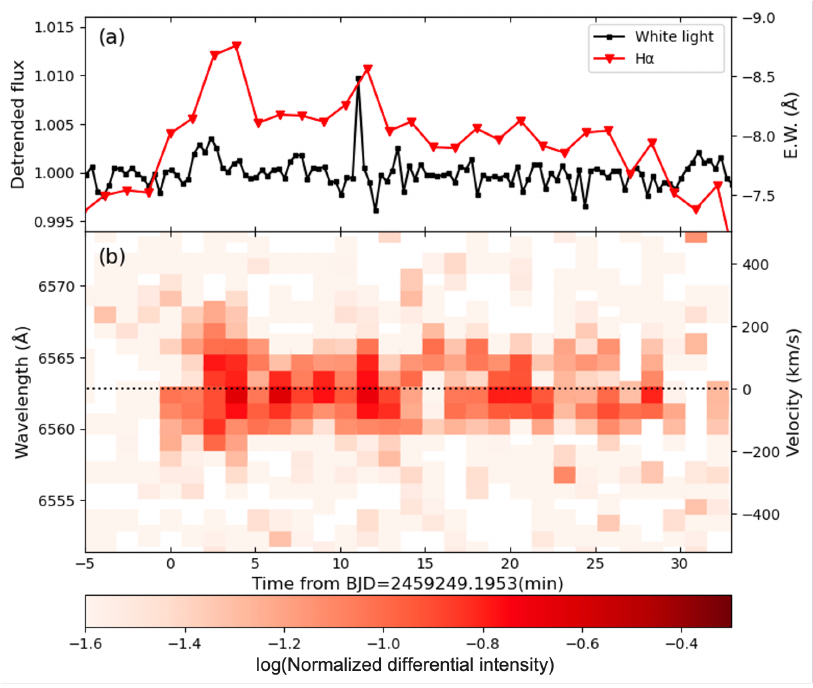}  
\figsetgrpnote{The same as Figure, but for the events of Y24.} 
\figsetgrpend

\figsetgrpstart
\figsetgrpnum{16.14} 
\figsetgrptitle{Light curves and dynamic spectrum in H$\alpha$ line for Y25 and 26} 
\figsetplot{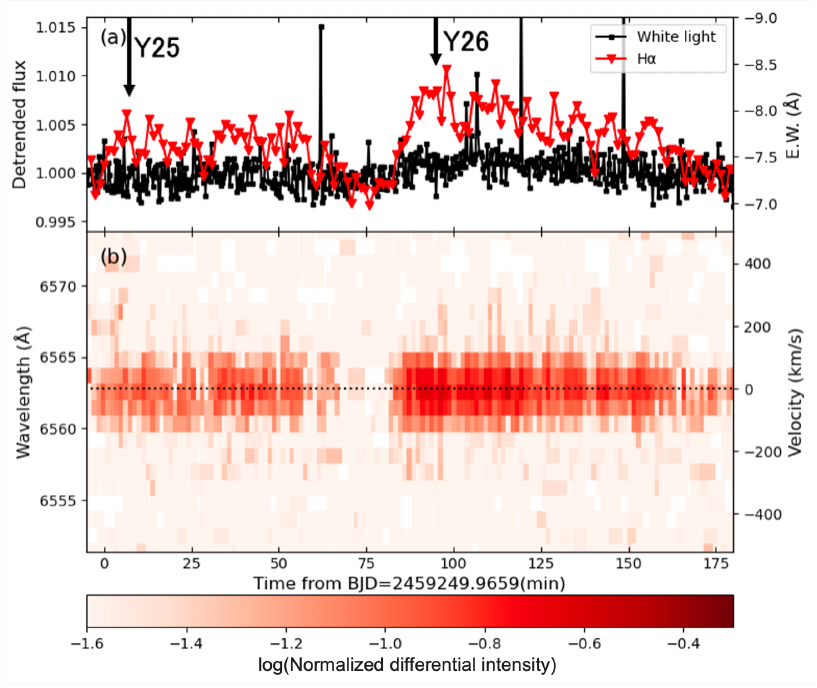}  
\figsetgrpnote{The same as Figure, but for the events of Y25 and 26.} 
\figsetgrpend

\figsetgrpstart
\figsetgrpnum{16.15} 
\figsetgrptitle{Light curves and dynamic spectrum in H$\alpha$ line for Y27} 
\figsetplot{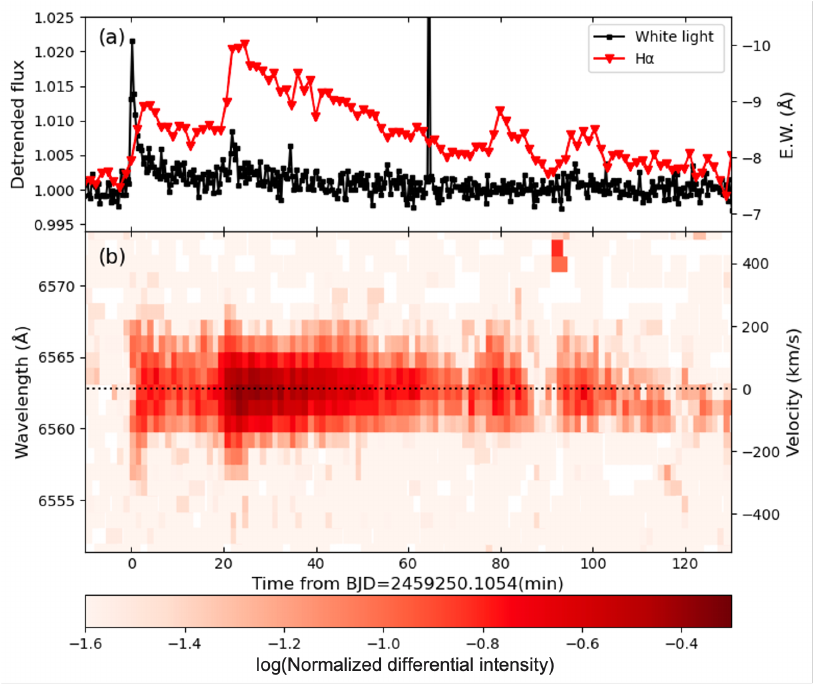}  
\figsetgrpnote{The same as Figure, but for the events of Y27.}  
\figsetgrpend

\figsetend

\clearpage

\bibliography{main}{}
\bibliographystyle{aasjournal}

\end{document}